\documentclass[aps,prfluids,12pt,onecolumn,amsmath,amssymb]{revtex4-2}

\usepackage{graphicx}
\usepackage{dcolumn}
\usepackage{bm}
\usepackage{amsmath,amssymb,amsthm}
\usepackage{hyperref}
\usepackage{cleveref}
\usepackage{booktabs}
\usepackage{hyperref}
\hypersetup{colorlinks=true,linkcolor=black,urlcolor=black,citecolor=black}

\renewcommand{\u}{\boldsymbol{u}}
\newcommand{\pp}{\partial}
\newcommand{\dd}[2]{\frac{\pp #1}{\pp #2}}
\newcommand{\st}{\boldsymbol{s}}
\newcommand{\stS}{{\cal S}}
\newcommand{\ac}{\boldsymbol{a}}
\newcommand{\acS}{{\cal A}}
\newcommand{\re}{r}
\newcommand{\trans}{p}
\newcommand{\params}{\boldsymbol{\vartheta}}

\begin{document}
\title{Drag Reduction in Flows Past 2D and 3D Circular Cylinders Through Deep Reinforcement Learning}

\newcommand{\cselab}{Computational Science and Engineering Laboratory, ETH Z\"{u}rich, CH-8092, Switzerland.}

\newcommand{\harvard}{Computational Science and Engineering Laboratory, Harvard University, Cambridge, MA 02138, USA.}

\author{Michail Chatzimanolakis${}^{1,2}$}
\author{Pascal Weber${}^{1,2}$}
\author{Petros Koumoutsakos${}^{2}$}
\thanks{corresponding Author: petros@seas.harvard.edu}
\affiliation{${}^1$\cselab}
\affiliation{${}^2$\harvard}
\date{\today}

\begin{abstract}
We investigate drag reduction mechanisms in flows past two- and three-dimensional cylinders controlled by surface actuators using deep reinforcement learning. We investigate 2D and 3D flows at Reynolds numbers up to $8'000$ and $4'000$, respectively. The learning agents are trained in planar flows at various Reynolds numbers, with constraints on the available actuation energy. The discovered actuation policies exhibit intriguing generalization capabilities, enabling open-loop control even for Reynolds numbers beyond their training range. Remarkably, the discovered two-dimensional controls, inducing delayed separation, are transferable to three-dimensional cylinder flows. We examine the trade-offs between drag reduction and energy input while discussing the associated mechanisms. The present work paves the way for control of unsteady separated flows via interpretable control strategies discovered through deep reinforcement learning.
\end{abstract}

\maketitle

\section{Introduction}
The identification and exploitation of drag reduction mechanisms for bluff body flows is at the core of  aircraft and ship design with direct impact on their power consumption and emissions. Flows  past circular cylinders have  long served as a classical prototype for such  drag reduction studies. We distinguish passive  control methodologies propose modifications to the body's surface, such as wall protrusions or surface roughness~\cite{Sirovich1997}, and active methods that involve surface  actuators, such tangential actuators, plasma, or mass transpiration ~\cite{Sosa2009,Milano2002,Wang2021}.
Automated discovery of control strategies for cylinder flows was first introduced to  optimize tangential actuators  on cylinder surfaces, exhibiting up to $50\%$ reduction in drag at a Reynolds number of 500~\cite{Milano2002}. Further investigations extended these findings to three-dimensional settings, achieving a $40\%$ reduction in drag~\cite{Milano2002}. \textcolor{black}{In \cite{Guinness2021} a series of simulations investigated passive drag reduction through partial leeward porous coatings on a cylinder's surface. The authors from \cite{Lee2004} experimented with the installation of a small control rod upstream of a cylinder, which affected its wake and drag coefficient. Another series of experiments were performed by \cite{Schulmeister2017}, who achieved active drag reduction for a cylinder by adding two small, counter-rotating cylinders near its surface.}

\textcolor{black}{
Reinforcement Learning (RL) has been introduced in fluid dynamics over the last decade first for developing control strategies for synchronizing multiple hydrodynamically interacting swimmers ~\cite{gazzola2014reinforcement,gazzola2016learning,Novati2017,Verma2018,mandralis2021learning}.The scope of these  applications has been extended in several directions including effective navigation in vortical flows~\cite{Gunnarson2021}, the semi-supervised discovery of subgrid-scale models in turbulent flows~\cite{novati2021a,Bae2022,Zhou2023} .
}

Reinforcement learning studies have also been introduced in flow control of prototypical flows in experiments and simulations \cite{rabault2020deep,Fan2020,Sonoda2023,vignon2023recent}.
Recent studies have explored drag reduction for planar flows past a cylinder using RL. \textcolor{black}{\cite{Tokarev2020} simulated an oscillating cylinder at a relatively low Reynolds number of $100$ and was able to achieve a drag reduction of $14\%$ to $16\%$, by controlling its angular velocity with RL.} For Reynolds numbers ranging from $100$ to $2000$, similar studies achieved reductions ranging from $17\%$, to $38\%$ ~\cite{Tang2020,Varela2022}, this time for a fixed cylinder. \textcolor{black}{RL has also been applied to experimental studies~\cite{Fan2020}. The experiments were performed for three circular cylinders to find a policy to reduce drag or increase the system power gain efficiency. They complemented their experiments with simulations that found similar policies. The authors reported  a wall-clock time of more than three weeks for their training to complete, exemplifying the high computational cost of these tasks.}

In this work we explore drag reduction mechanisms for the flow past a circular cylinder for Reynolds numbers up to $8000$ using Deep Reinforcement Learning (DRL). We develop a control policy and examine its physical characteristics and its generalisation capabilities not only to different Reynolds numbers but also from two to three-dimensional flows. The flow features are captured through an efficient implementation of Adaptive Mesh Refinement \cite{chatzimanolakis2022b}. The  computational cost associated with the stochastic search involved in DRL is alleviated by a parallel implementation of V-RACER with Remember and Forget for Experience Replay (ReF-ER)~\cite{Novati2019} in Korali~\cite{martin2021korali}, which allows parallel training with direct numerical simulations at a manageable cost.

\textcolor{black}
{This paper is structured as follows: In~\cref{sec:Governing Equations}, we introduce the governing equations and numerical method used for our Direct Numerical Simulations. The Reinforcement Learning method is discussed in~\cref{sec:Reinforcement Learning}. Building on that background, we formulate drag reduction as a Reinforcement Learning problem in~\cref{sec:RL for Drag Reduction}. Our results, presented in~\cref{sec:Results}, are organized as follows:First, we introduce the experimental setup and examine the mechanism for drag reduction on one example in~\cref{sec:mechanism}. Then, in~\cref{sec:effect velocity}, we analyze the effect of the actuation velocity on the results. The transferability of the learned policy to different Reynolds numbers is discussed in~\cref{sec:effect reynolds}, and we further investigate the influence of the two factors on the actions taken in~\cref{sec:discussion actions}.We explore extending the results to 3D in~\cref{sec:three dimensions} and  conclude in~\cref{sec:conclusions}.
}

\subsection{Direct Numerical Simulations} \label{sec:Governing Equations}
We perform two- and three-dimensional Direct Numerical Simulations (DNS) of the flow past a cylinder, by solving the incompressible Navier-Stokes equations
\begin{equation}\label{eq:Incompressible-Navier-Stokes}
\begin{split}
&\mathbf{\nabla}\cdot\u=0\,,\\
&\dd{\u}{t}+(\u \cdot \mathbf{\nabla})\u=-\frac{1}{\rho}\nabla p+\nu\mathbf{\nabla}^2 \u,
\end{split}
\end{equation}
where $\u,\rho,p$ and $\nu$ are the fluid velocity, density, pressure and kinematic viscosity. The no-slip boundary condition is enforced on the cylinder surface with a prescribed velocity $\u^{s}$ through the penalisation approach \cite{Angot1999,Ueda2021, Rossinelli2015}, which augments the Navier-Stokes equations with a penalty term $\lambda \chi (\u^{s}-\u)$. Here $\lambda \in \mathbb{R}$ is the penalisation coefficient and $\chi$ is the characteristic function that takes values $\chi=1$ inside the cylinder and $\chi=0$ outside. The simulations are performed with the \textsc{CubismAMR} software, an adaptive version of the \textsc{Cubism} library, which partitions the simulation domain into cubic blocks of uniform resolution that are distributed to multiple compute nodes for cache-optimised parallelism~\cite{Rossinelli2013}. \textsc{CubismAMR} organizes these blocks in an octree data structure (for three-dimensional simulations) or a quadtree data structure (for two-dimensional simulations), allowing for Adaptive Mesh Refinement in different regions. We refer to \cite{Chatzimanolakis2022a} and \cite{chatzimanolakis2022b} for details on the implemented numerical scheme and code validation results.

\subsection{Reinforcement Learning} \label{sec:Reinforcement Learning}
RL algorithms solve Markov Decision Processes (MDPs), which are defined by the tuple $(\stS,\;\acS,\;\re,\;\trans)$ consisting of a state-space $\stS$, an action-space $\acS$, a function $\re:\stS\times\stS\times\acS\to\mathbb{R}$ which is the reward of transitioning to state $\st' \in\stS$ from state $\st \in\stS$ by taking action $\ac\in\acS$, and an unknown, stochastic transition map $\trans(\st'|\ac,\st)$, which is the probability of transitioning to $\st'$ from $\st$ by taking action $\ac$.

On the MDP, we define a stochastic policy via a probability distribution ${\pi}(\ac|\st)$, which allows sampling an action for a given state. The goal of RL is to find the optimal policy
\begin{equation}
{\pi}^\star=\arg\max\limits_\pi V^{\pi}(\st)\,,\quad \forall \st\in 
\stS\,.
\end{equation}
that maximizes the state-value function, defined as
\begin{equation}
{V}^{\pi}(\st)=\mathbb{E}_{\trans,\pi}\left[\sum\limits_{i=0}^{N_i-1} \gamma^i \re(\st_i,\st_{i+1},\ac_i)|\st_0=\st\right]\,,
\end{equation}
where $\gamma^i\in [0,1)$ is known as the “discount factor" and $N_i$ are the total number of transitions between states.

The optimal policy is computed based on interactions of an RL agent with the environment. At every step $i$, the agent chooses an action $\ac_i$ based on the observation of the state $\st_{i}$ from the environment. The environment then transitions to a new state $\st_{i+1}$ and returns a reward $\re(\st_i,\st_{i+1},\ac_i)$. In off-policy methods, transitions are collected in a Replay Memory and an approximation to the optimal policy ${\pi}(\ac|\st;\params)$ with parameters $\params$ is learned. In actor-critic methods, a value function $V(\st;\params)$ approximation is learned as well. State-of-the-art DRL employs neural networks $\operatorname{NN}(\st;\params)$ as universal function approximators, where the weights $\params$ of the neural network are typically optimized using stochastic gradient descent. For the present work, we use V-RACER with ReF-ER~\cite{Novati2019} implemented in Korali~\cite{martin2021korali}; this is an off-policy actor-critic DRL method, proven successful in several scientific applications ~\cite{Novati2021,Bae2022} and recently generalized to multiple RL agents~\cite{Weber2022}.

\section{Reinforcement Learning for Drag Reduction}\label{sec:RL for Drag Reduction}
The effective deployment of  RL requires an appropriate choice of states, actions, and reward function.  Here we deploy a RL agent that interacts with the environment at discrete times $t_i=t_0 + i \Delta t$, for $i=0,\dots,N_i-1$, where $N_i$ is the total number of actions taken before a set of interactions, also referred to as an episode, terminates. The times at which the actions are taken are equally spaced in time, with spacing $\Delta t$, and the agent starts the interaction after a transient time $t_0$.  We set $\Delta t = 0.1 U/D$, where $U$ is the cylinder velocity and $D$ its diameter.

\paragraph{Actions:}
We deploy $N_a=8$ uniformly distributed mass transpiration actuators on the cylinder surface~\cite{Milano2002,Tang2020}, each with a time-dependent strength $a_j^{t_i}\in[-1,1]~,~j=0,\dots,N_a-1$. Actuator $j$ imposes a radial velocity on the surface of the cylinder:
\begin{equation}\label{eq:action}
v^r_j(\theta) = cUa_j^{t_i} \cos{\left( \frac{\pi (\theta-\theta_j)}{\theta_a^j}\right)}\,, |\theta-\theta_j| \le \theta_a^j/2\,,
\end{equation}
where $c$ is a constant that varies during training, $\theta_a^j$ is the angle that corresponds to the arc length over which actuator $j$ is active, $\theta_j = j \frac{\pi}{4}$ is the angle on the cylinder surface where each actuator is centered. One set of actions $\ac_i$ consists of picking the actuator strengths under the constraint that their mean value $\sum_{j=1}^N a_j^{t_i}/N_a$ is zero, ensuring a zero total mass flux caused by actuation.

\paragraph{State:}
We observe the cylinder lift and drag coefficients as well as the pressure and vorticity on its surface in $N_s=16$ uniformly placed sensors. For each of the sensors, pressure and vorticity are averaged over the cylinder surface in an area covered by an arc-length of $10$ degrees. We also include the Reynolds number and $c$ value from \cref{eq:action} in our state representation, resulting in a $36-$dimensional state.

\paragraph{Reward:} The reward function entails the averaged total drag and a penalty term for the actuation strengths, expressed as:
\begin{equation}
    r(\st_i,\st_{i+1},\ac_{i}) = -\frac{1}{\Delta t} \int_{t_{i}}^{t_{i+1}} C_D(t)\;\mathrm{d}t - \frac{w}{N_a}\sqrt{\sum_{j=0}^{N_a}(a^{t_i}_j)^2}\,,
\end{equation}
where $C_D$ denotes the cylinder drag coefficient. 
The first term implies that the the agent minimizes the mean drag. The second term is a regularization term with coefficient $w$, that penalizes strong actions and therefore balances the trade-off between drag reduction and energy required for the actuation.

\section{Results}\label{sec:Results}
\begin{figure}
    \centering
    \includegraphics[trim=450 400 0 400, clip, width=\linewidth]{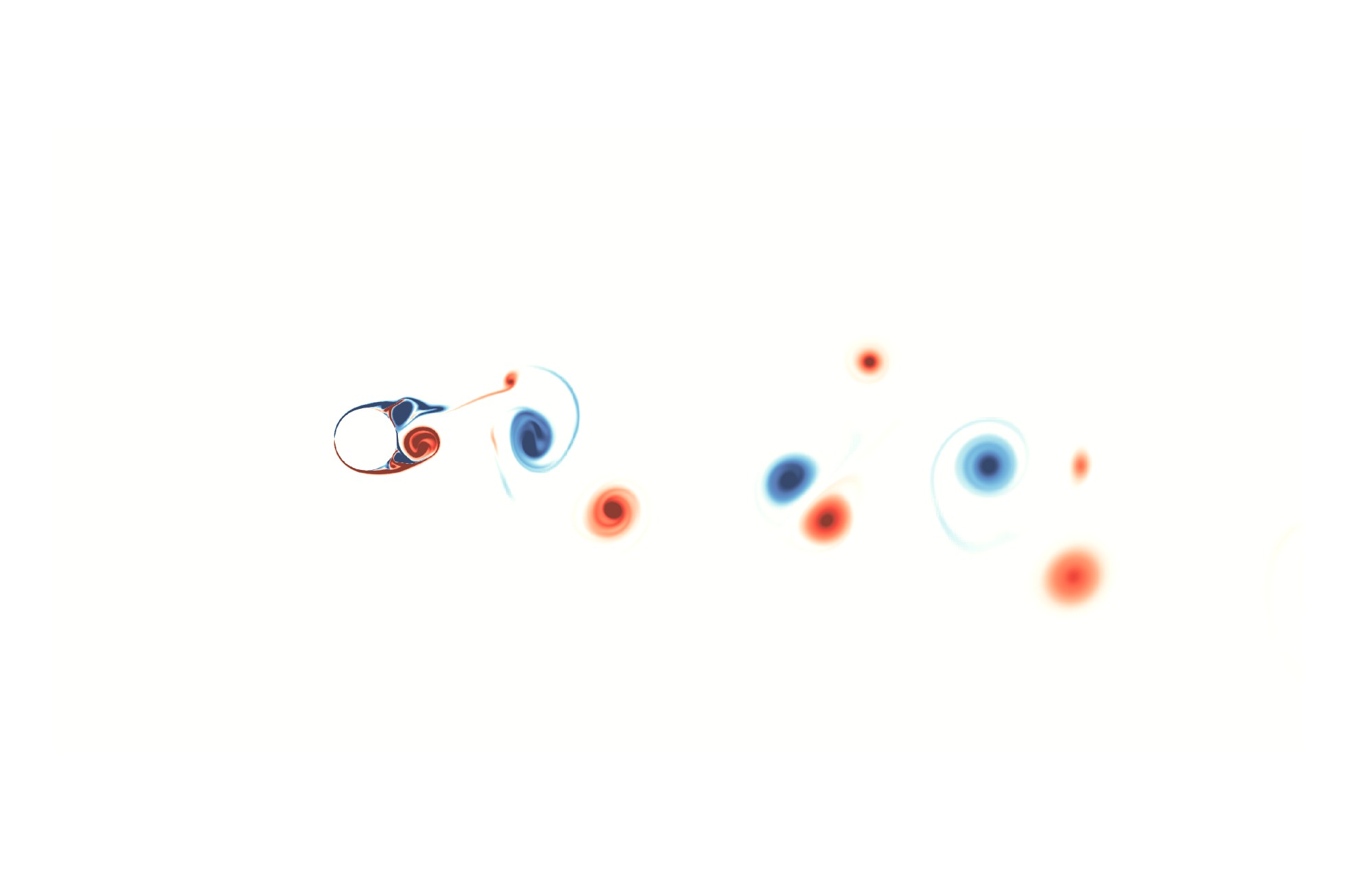}
    \includegraphics[width=.49\linewidth]{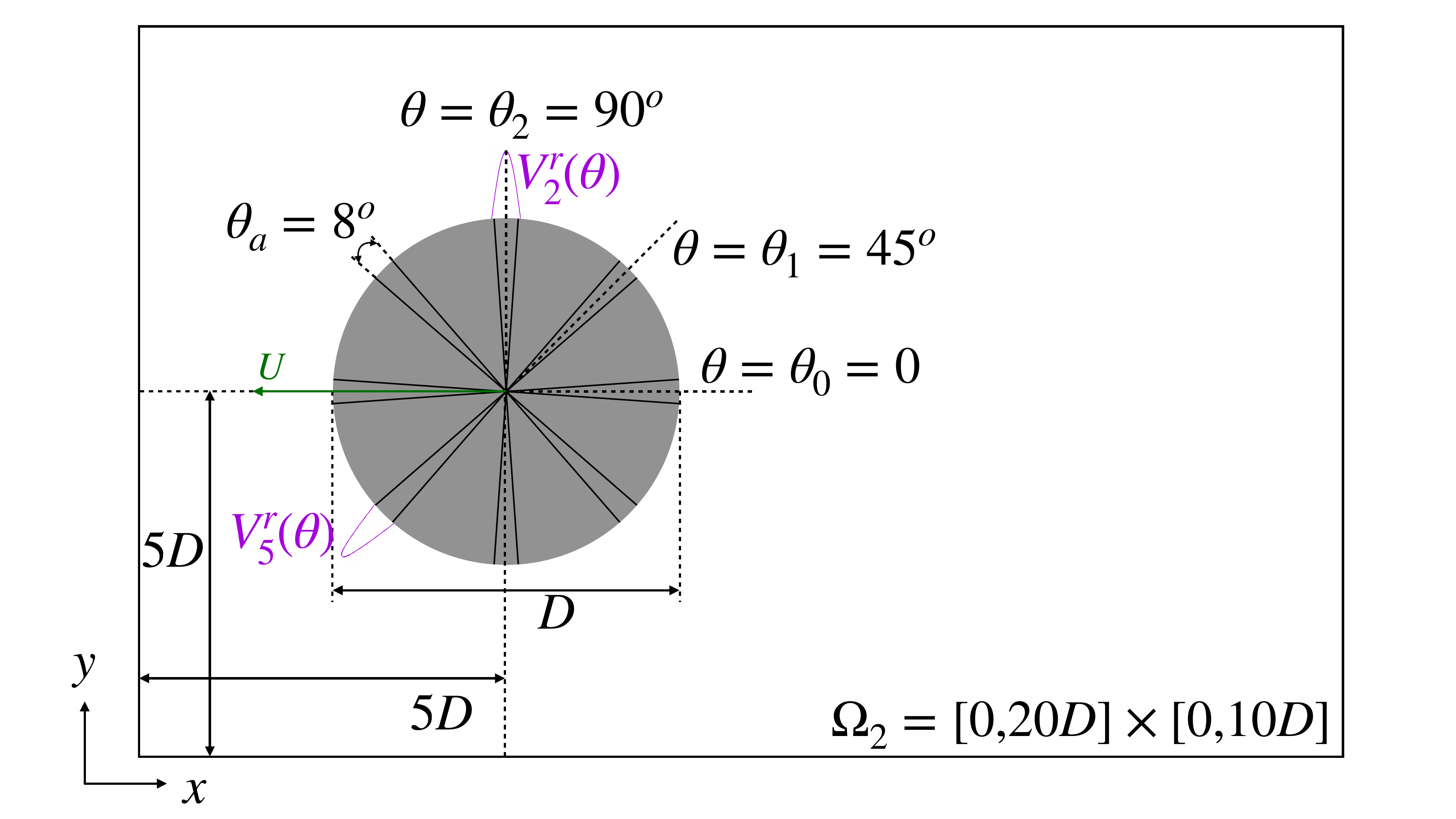}
    \includegraphics[width=.49\linewidth]{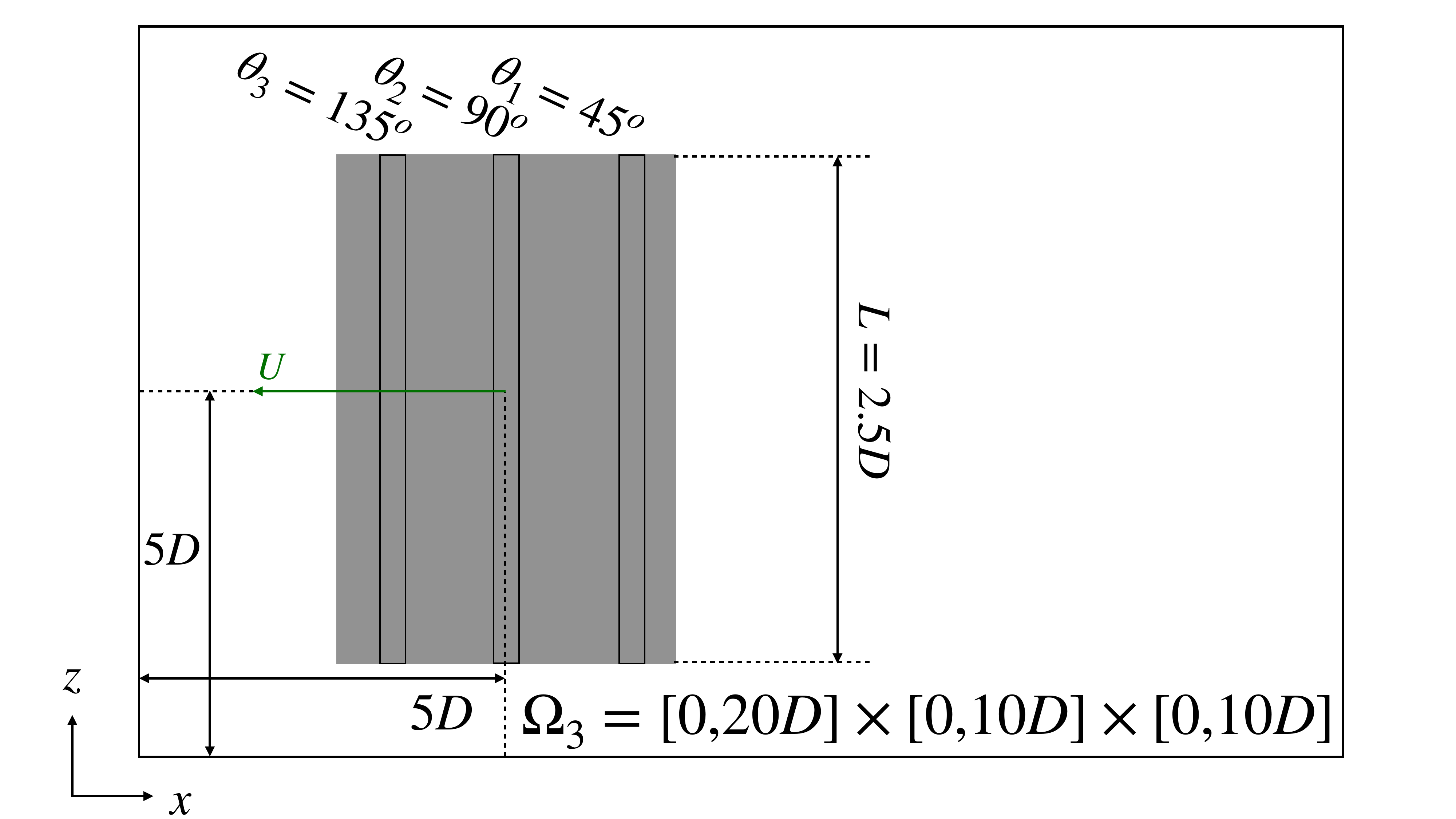}
    \caption{Illustration of the RL setup. Top: snapshot of the vorticity field at $T=200$ for the two-dimensional flow at $Re=4000$. Bottom left: Sketch of the actuators and 2D simulation domain. Bottom right: sketch of the actuators and 3D simulation domain, top view.}
    \label{fig:simulation_setup}
\end{figure}

We perform two and three-dimensional simulations of flows past circular cylinders. The two-dimensional simulations are performed in a rectangular domain $\Omega_2 = [0,20D]\times[0,10D]$, with a cylinder of diameter $D$ placed at $(5D,5D)$. The three-dimensional simulations use a rectangular domain $\Omega_3 = [0,20D]\times[0,10D]\times[0,10D]$, with a cylinder of diameter $D$ and length $L=2.5D$ placed at $(5D,5D,5D)$. In both cases, the cylinder is impulsively set into motion with velocity $U$. \Cref{fig:simulation_setup} shows the aforementioned setup. Adaptive mesh refinement  takes place according to the magnitude of the vorticity field and the finest resolution depends on the minimum grid spacing allowed (denoted by $h$). We find that decreasing  the value of $h$ below $D/200$ yields a smaller than $2\%$ change in the computed mean drag, and thus choose to use this value for our simulations. Non-dimensional time is scaled as $T=tU/D$ while the timestep is determined according to the Courant–Friedrichs–Lewy (CFL) condition, with a Courant number of $0.5$. Commencement of vortex-shedding is accelerated by adding a small rotation of the cylinder along the $z-$ direction. For $0.25<T<0.5$ the $z-$ component of its angular velocity is
\begin{equation}
    \omega_z = \frac{0.04U}{D} \sin\left(8\pi T\right) \,.
\end{equation}
We train the RL agent in 2D simulations with a Reynolds number randomly sampled from $Re=1000,2000$ and $4000$. For each episode, the maximum actuation velocity is determined by sampling $c$, \cref{eq:action}, in the interval $[0.05,0.15]$. We examine two policies $\pi$ and $\pi^w$. The first policy does not employ a regularizer ($w=0$) whereas the second policy does (we set $w=0.1$). Each episode consists of a simulation where actuation starts at $T=200$, when the wake of the cylinder is developed and vortex shedding has commenced (see also \cref{fig:simulation_setup} for a snapshot of the vorticity field) and ends at $T=250$. Each training was run on $32$ compute nodes, each equipped with two AMD EPYC 7763 of $64$ cores and lasts $12$ hours. As can be seen in the top left plot of~\cref{fig:AllPolicy01}, this allows simulating approximately $4000$ episodes and achieves a converged policy approximately after episode $3000$. Note that the resolution used during training used a minimum grid spacing of $2h$ as a compromise between time-to-solution and accuracy of each episode simulated. After training, the found policies are tested in a fully resolved simulation for different Reynolds numbers and maximum actuation velocities (determined by the value of $c$). \Cref{tab:summary-2d} compares the mean drag for the uncontrolled case $T\in[100,200]$ and for the controlled case where $T\in[200,300]$.

\begin{table}
\centering
\caption{Summary of the two-dimensional simulation results. Each row shows the mean drag coefficient without actuation and the mean drag coefficient for three cases with different values for the maximum actuation velocity $c$. Results are shown for both policies $\pi$ and $\pi^w$. The percentage written in parentheses indicates the drag reduction.}\label{tab:summary-2d}
\begin{ruledtabular}
\begin{tabular}{*{2}{l}|*{3}{l}|*{3}{l}}
\multicolumn{2}{c}{} & \multicolumn{3}{c}{Policy $\pi$ (no regularizer weight)} & \multicolumn{3}{c}{Policy $\pi^w$ (regularizer weight $w=0.1$)} \\
\toprule
   Re  & $C_D$\;\;\;\; & $c=0.05$ & $c=0.10$ & $c=0.15$\;\;\;\;\;\;\;\;  &  $c=0.05$ & $c=0.10$ & $c=0.15$ \\ \midrule
 500 & 1.52 & 1.42 ($ 6\%$) &  1.33 ($12\%$) &  1.25 ($18\%$) & 1.43 ($ 6\%$) &  1.34 ($12\%$) &  1.26 ($17\%$) \\ 
1000 & 1.61 & 1.45 ($10\%$) &  1.31 ($19\%$) &  1.20 ($26\%$) & 1.45 ($ 10\%$) &  1.32 ($18\%$) &  1.21 ($25\%$) \\ 
2000 & 1.75 & 1.52 ($14\%$) &  1.28 ($27\%$) &  1.14 ($35\%$) & 1.53 ($13\%$) &  1.30 ($26\%$) &  1.16 ($34\%$)\\ 
4000 & 1.75 & 1.70 ($ 3\%$) &  1.35 ($23\%$) &  1.18 ($33\%$) & 1.70 ($ 3\%$) &  1.42 ($19\%$) &  1.20 ($32\%$) \\ 
8000 & 1.92 & 1.91 ($ 1\%$) &  1.67 ($13\%$) &  1.29 ($33\%$) & 1.67 ($13\%$) &  1.74 ($ 10\%$) &  1.33 ($31\%$) \\
\end{tabular}
\end{ruledtabular}
\end{table}

\subsection{Mechanism for Drag Reduction}\label{sec:mechanism}
\begin{figure}
    \centering
    \includegraphics[width=\linewidth]{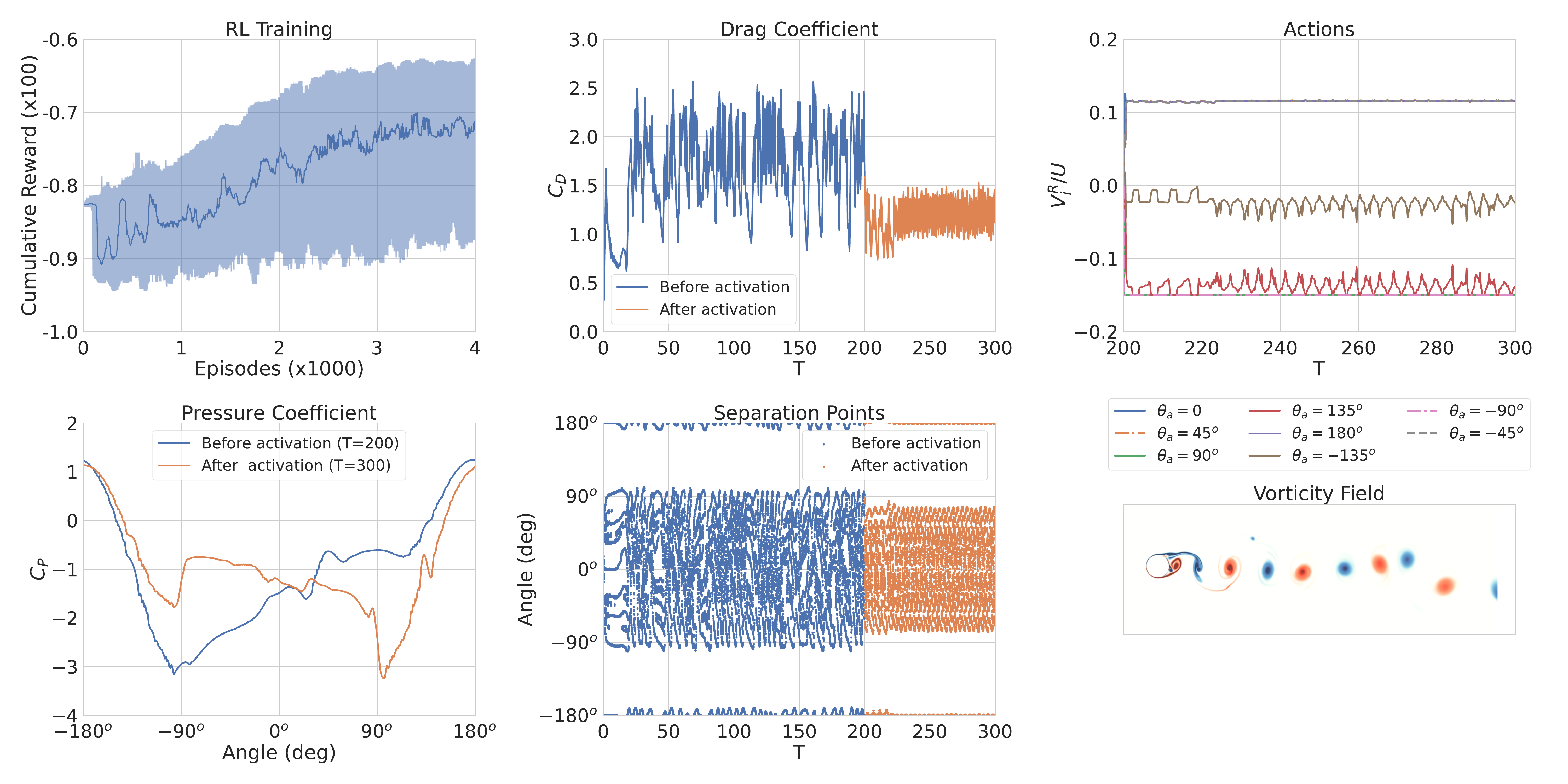}
    \caption{Results for policy $\pi$ (no regularizer weight) for a two-dimensional flow past a cylinder at $Re=4000$, with maximum actuation velocity up to $15\%$ of the cylinder velocity. Top left: Cumulative reward as a function of episodes simulated, plotted with a $95\%$ confidence interval. Top center: Drag coefficient as a function of time for the converged policy. Top right: Actions as fraction of cylinder velocity as a function of time. Bottom left: Cylinder pressure coefficient at two time instances, before and after actuators are activated. Bottom center: Angle of the points with zero shear (separation points) on cylinder surface. Bottom right: Vorticity field at $T=300$.}
    \label{fig:AllPolicy01}
\end{figure}

In order to understand the mechanism for drag reduction, we present the results for $Re=4000$ and $c=0.15$ in~\cref{fig:AllPolicy01}. In the top left plot, the drag coefficient time series before and after the actuators are activated is presented; drag is decreased by almost $33\%$ (see \cref{tab:summary-2d}). The fluid velocity imposed by each actuator expressed as a fraction of the cylinder velocity is displayed in the top right panel. Positive values correspond to blowing and negative values to suction. The actuators at the front half of the cylinder ($\pi/2 \le |\theta_a| \le \pi$) suction fluid, while the other actuators blow fluid, which helps the boundary layer remain attached for longer. The vorticity field with actuation (bottom right plot of \cref{fig:AllPolicy01}) at $T=300$ can be compared with the initial condition at $T=200$ shown in~\cref{fig:simulation_setup}. The comparison reveals that the cylinder wake becomes narrower and more symmetric. The width of the wake is closely associated with the location of the flow separation point on the cylinder surface, where early separation can cause wider wakes and increased drag~\cite{chatzimanolakis2022b}. 
The policy successfully delays separation, allowing the flow to remain attached for a longer duration and resulting in reduced drag. This delay of the separation is quantified in the bottom left and center plots of~\cref{fig:AllPolicy01}. The first plot shows the pressure coefficient on the cylinder surface at $T=200$ and $T=300$, corresponding to time instances before and after the activation of actuators. Regions of separated flow are characterized by flat pressure profiles, which significantly contribute to drag increases. After the actuators are activated, the regions of constant pressure become visibly smaller, indicating that the flow remains attached for a longer duration. The center plot shows the polar angle of the points of zero vorticity on the cylinder surface over time, which are indicative of flow separation. The found policy effectively moves the separation angle towards the back of the cylinder, reducing the maximum separation angle (at the top part of the cylinder) from approximately 90 degrees to around $\pm$75 degrees.
Overall, the results from policy $\pi$ demonstrate the effectiveness of the implemented drag reduction strategy, as evidenced by the significant reduction in drag, improved wake symmetry, and delayed separation.

\subsection{Effect of Actuation Velocities}\label{sec:effect velocity}
%
\begin{figure*}[!b]
    \centering
    \includegraphics[width=\linewidth]{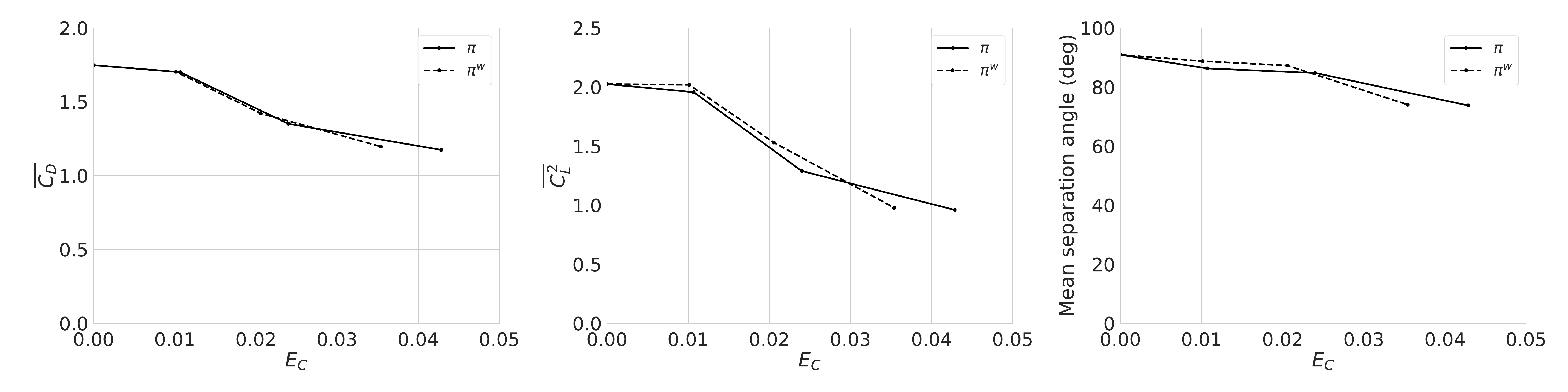}
     \includegraphics[width=\linewidth]{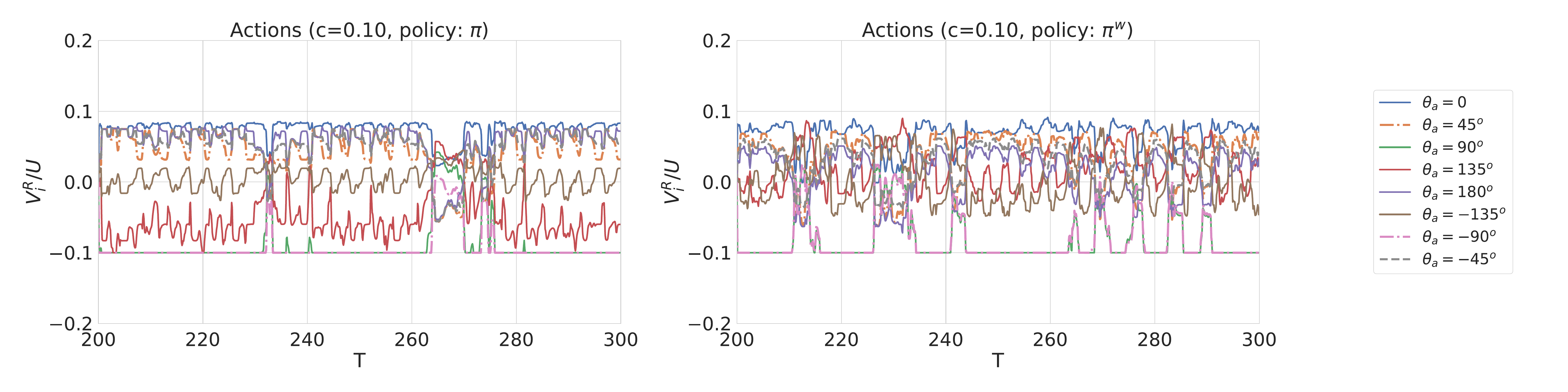}
       \caption{Summary of results when a regularizer is introduced into the reward function, for $Re=4000$. Top row: mean drag coefficient, mean squared lift coefficient and mean flow separation angle, as functions of the energy cost for policy $\pi$ (no regularizer) and policy $\pi^w$ (with regularizer). Bottom: comparison of actions taken by the two policies, expressed as a fraction of the cylinder velocity, for $c=0.10$.
    }
    \label{fig:energycost4000}
\end{figure*}

We examine the effect of the actuation strength and of introducing a regularizer preferring low actuation to the reward function. The actuation velocity effectively influences the amount of energy injected into the system. To quantify this added energy, we introduce the energy cost metric, denoted as $E_c$. It is defined as follows:
\begin{equation}
    E_c = \frac{1}{(\tau_2-\tau_1)N_a}\int_{\tau_1}^{\tau_2}
    \sqrt{ \sum_{j=0}^{N_a-1} (a_j^{t_i})^2} \mathrm{d}t\ \\,
\end{equation}
where $\tau_1=200T$ corresponds to the start of actuator activation and $\tau_2=300T$ to the end time of each of our simulations. We also define $E_c = 0$ for simulations with inactive actuators. 
This metric provides a measure of the overall cost associated with reducing drag by activating the actuators within a given time interval. \Cref{fig:energycost4000}, shows the mean drag coefficient, mean squared lift coefficient, and mean separation angle as functions of the energy cost. Notably, as the energy cost increases, we observe more substantial reductions in drag. These reductions are achieved by generating narrower and more symmetric wakes. Consequently, the mean squared lift coefficient and mean separation angle decrease with increasing energy cost. Note that an energy cost of zero corresponds to the baseline case, before the actuators were activated. As anticipated, smaller actuation velocities result in less drag reduction, refer also to \cref{tab:summary-2d}. It is worth noting that the inclusion of a regularizer in the reward function yields comparable drag reductions to the case without a regularizer. However, a significant difference lies in the energy cost, particularly when higher actuation velocities ($c=0.10$ or $c=0.15$) are permitted. In both instances, the energy cost is approximately $25\%$ lower.

\begin{figure*}
    \centering
    \includegraphics[width=\linewidth]{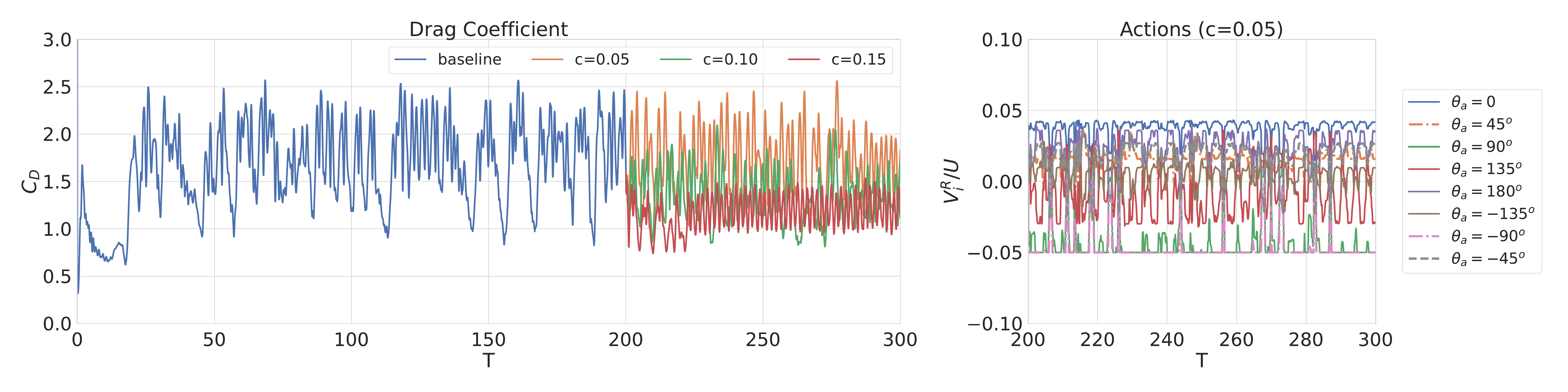}
    \includegraphics[width=\linewidth]{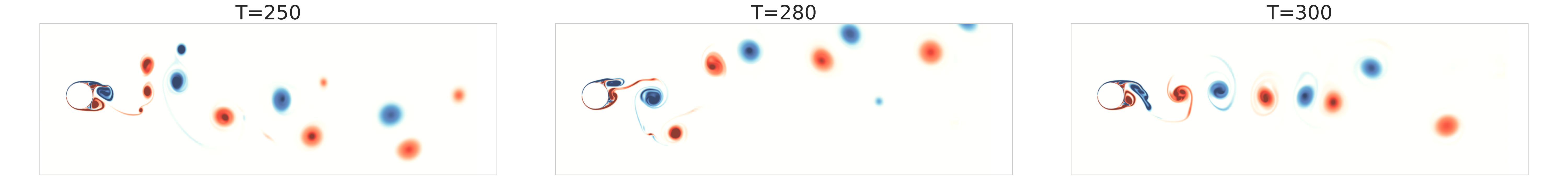}
    \caption{Summary of results for policy $\pi$ with varying maximum actuation velocities (varying $c$) and a fixed Reynolds number of $4000$. Top left: Drag coefficient time series for different values of $c$.
    Top right: Actions time series for $c=0.05$, expressed as fraction of cylinder velocity. Bottom: vorticity field at several time instances, for $c=0.05$.}
     \label{fig:weights}
\end{figure*}

To understand how varying the maximum actuation velocity affects our policy, we plot the drag coefficient for $Re=4000$ and different values of $c$, in \cref{fig:weights}. We see that for $c=0.05$ drag is not reduced until $t\approx 280T$. The vorticity field for $c=0.05$ is presented in \cref{fig:weights}, for various time instances. Like the stronger actuation case, the wake width eventually becomes narrower here. However, this does not happen until $t\approx 280T$, which coincides with the time instance during which drag is actually reduced. It is only after that time that the separation angle becomes acute, yielding a narrower wake and a decrease in drag of about $8\%$. When averaged throughout the whole simulation, the final decrease ends up being only $2.7\%$. Interestingly, for $c=0.05$, the actions taken by the RL agent exhibit greater variance over time compared to when $c=0.15$, as depicted in both \cref{fig:weights} and \cref{fig:AllPolicy01}. This variation can be attributed to the RL agent's efforts to control the wake of the cylinder, aiming to make it narrower and more symmetrical. For instance, we observe that the actuator located at $\theta_a=-90^o$ is occasionally turned off (the actuation strength briefly reaches zero, as indicated by the pink curve in \cref{fig:weights}). This typically occurs when vortices are shed in the positive $y$ direction, such as at $t\approx 280T$. A similar behavior is observed when vortices are shed in the negative $y$ direction, concerning the actuator located at $\theta_a=90^o$. Switching off or momentarily reducing the strength of these two actuators appears to be crucial in controlling the direction of vortices shed by the cylinder. Ultimately, successful control leads to a narrower wake and reduced drag. 

At the bottom of \cref{fig:weights}, we visualize the impact of introducing a regularizer on the actions taken by the RL agent for $c=0.1$. On the left, we display the actions taken by policy $\pi$, while on the right, we present the actions taken by $\pi^w$. Notably, we observe intermittent deactivation of the actuators positioned at $\theta_a = \pm 90^o$. As discussed earlier, strategic deactivation of these influential actuators at opportune moments enhances wake symmetry. In addition to that, the transition from $\pi$ to $\pi^w$ gives the RL agent the ability to identify safe time instances for deactivating these actuators without compromising drag reduction performance. Furthermore, when employing $\pi^w$, the actuator located at $\theta_a=135^o$ exhibits periodic oscillations around a small value, effectively reducing the energy cost. Conversely, when utilizing $\pi$, this particular actuator maintains a nonzero mean value without necessarily contributing to drag reduction significantly.

%

\subsection{Effect of Reynolds number}\label{sec:effect reynolds}
Here, we assess the effectiveness of our policy at varying Reynolds numbers. During training, we used $Re=1000,2000,4000$ and we extend this range by also testing $Re=500$ and $Re=8000$. 
\begin{figure*}
    \centering
    \includegraphics[width=\linewidth]{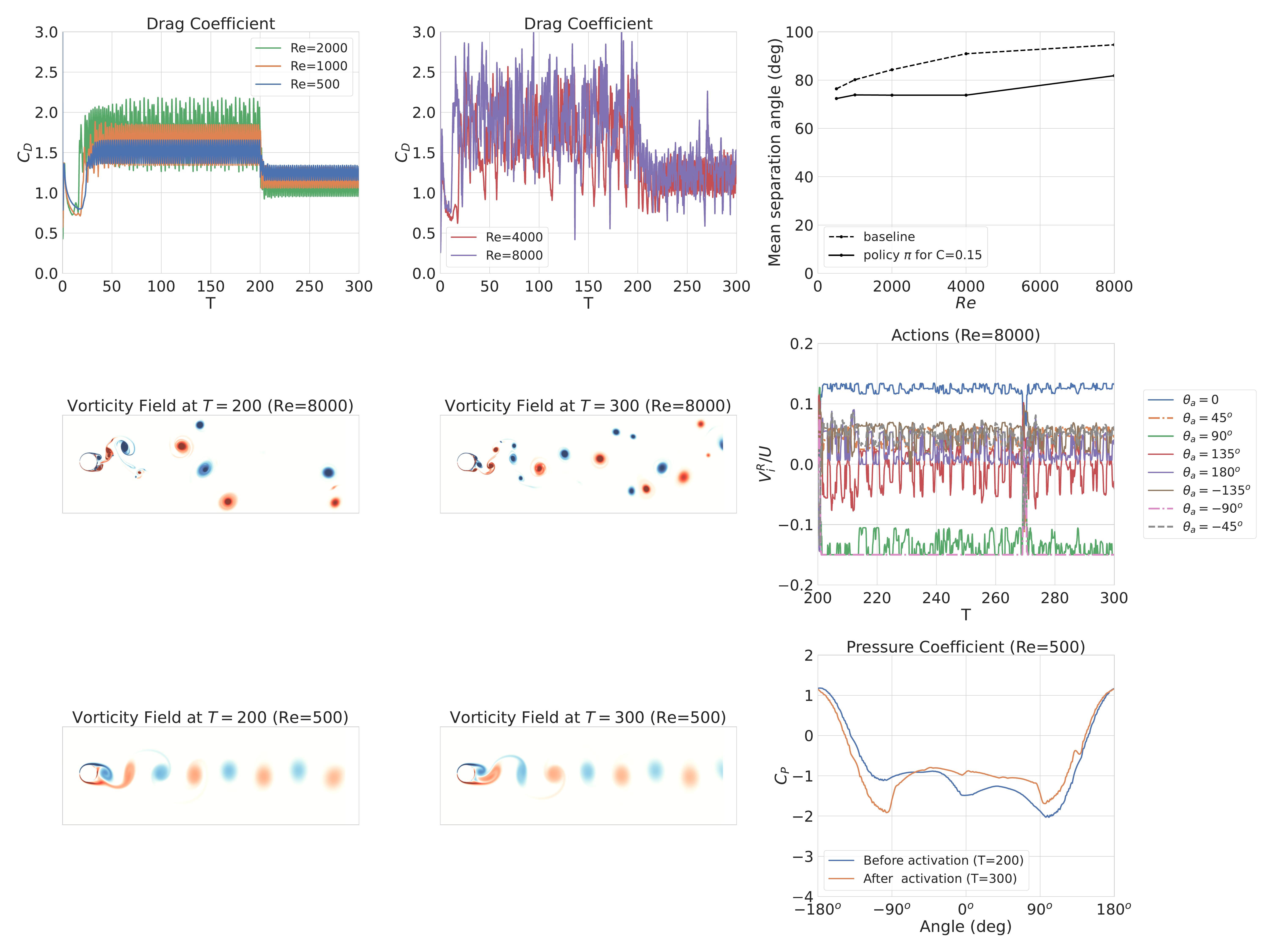}
    \caption{Summary of results for different Reynolds numbers and $c=0.15$. Top left and top center: drag coefficient time series. Top right: mean separation angle as a function of the Reynolds number for the uncontrolled case and a controlled case. Middle left and center: vorticity field at $Re=8000$ before ($T=200$) and after actuation ($T=300$). Middle right: Actions as fraction of cylinder velocity for Re=8000. Bottom left and center: vorticity field at $Re=500$ before ($T=200$) and after actuation ($T=300$). Bottom right: Cylinder pressure coefficient for $Re=500$ before and after actuators are activated.
}
    \label{fig:rey01}
\end{figure*}

For $c=0.15$, we observe a significant reduction in drag across all Reynolds numbers, as depicted in \cref{fig:rey01}. Similar to the case of $Re=4000$, the wake becomes narrower and flow separation is delayed, as evident by the comparison between the initial and final vorticity field for $Re=8000$ in the middle of \cref{fig:rey01}. The location of separation points and pressure coefficient profiles also show similar trends; the middle right of \cref{fig:rey01} shows how the mean separation point location varies with the Reynolds number for the uncontrolled scenario and when policy $\pi$ is applied for $c=0.15$. The displacement of the separation point is greater for the larger Reynolds numbers examined. For $Re=1000,2000,4000$ and $8000$, the mechanism for drag reduction is mostly the same for $c=0.15$. To elucidate the situation at lower Reynolds numbers, we examine the cylinder pressure coefficient. An example is shown in the bottom right plot of \cref{fig:rey01}. Although the size of constant pressure regions does not change significantly, the pressure values do, leading to the reduction in total drag. These subtle changes are not clearly visible in the vorticity field before and after actuation, which is shown at the bottom left and center of \cref{fig:rey01}. 

As the Reynolds number increases the actions taken by the RL agent demonstrate greater variance with time. This is evident is we compare the top left plot of \cref{fig:rey01} with the top left plot of \cref{fig:AllPolicy01}. We have established that a strategic deactivation of the actuators placed at $\theta_a = \pm 90^o$ can help with enforcing symmetry in the cylinder's wake. So far, this has only been necessary for smaller values of the actuation velocity ($c=0.05$ or $c=0.10$). This is not the case for $Re=8000$, where we see that large actuation values are not sufficient and need to be combined with the aforementioned periodic deactivation of the actuators, to manage to reduce drag and maintain more symmetric conditions in the wake, compared to the uncontrolled case.

When transitioning to the comparatively conservative scenario (with regard to action/actuator velocity magnitudes) characterized by $c=0.05$, our policy demonstrates notable success in mitigating drag, specifically for $Re=500, 1000$, and $2000$ (refer to \cref{tab:summary-2d}). In the case of $Re=4000$, it has been demonstrated that control eventually proves effective, resulting in an approximate $8\%$ reduction in drag. However, it is worth noting that the average reduction amounts to around $2\%$ due to the time required for the actions taken to manifest their full impact. At $Re=8000$, allowing actuation with only up to $5\%$ of the cylinder's velocity does not seem to suffice, to achieve a notable drag reduction. The introduction of a regularizer does not seem to alter results significantly, for $c=0.05$. This is valid across all Reynolds numbers, with the exception of $Re=8000$ where a $13\%$ reduction is observed. It is hard to pinpoint the reasons for this success, especially because it was not observed at $Re=4000$, which is a value used during training. We found that the drag coefficient for this case is somewhat reduced for $t\in[240,260]$ and $t\in[270,280]$. This reduction does not persist for later times, and we thus do not mainly attribute it to policy $\pi^w$ but conjecture that the chaotic nature of the flow also gives rise to some rare local drag minima.

\subsection{Discussion - action space}\label{sec:discussion actions}
\begin{figure}[ht]
    \centering
    \includegraphics[width=\linewidth]{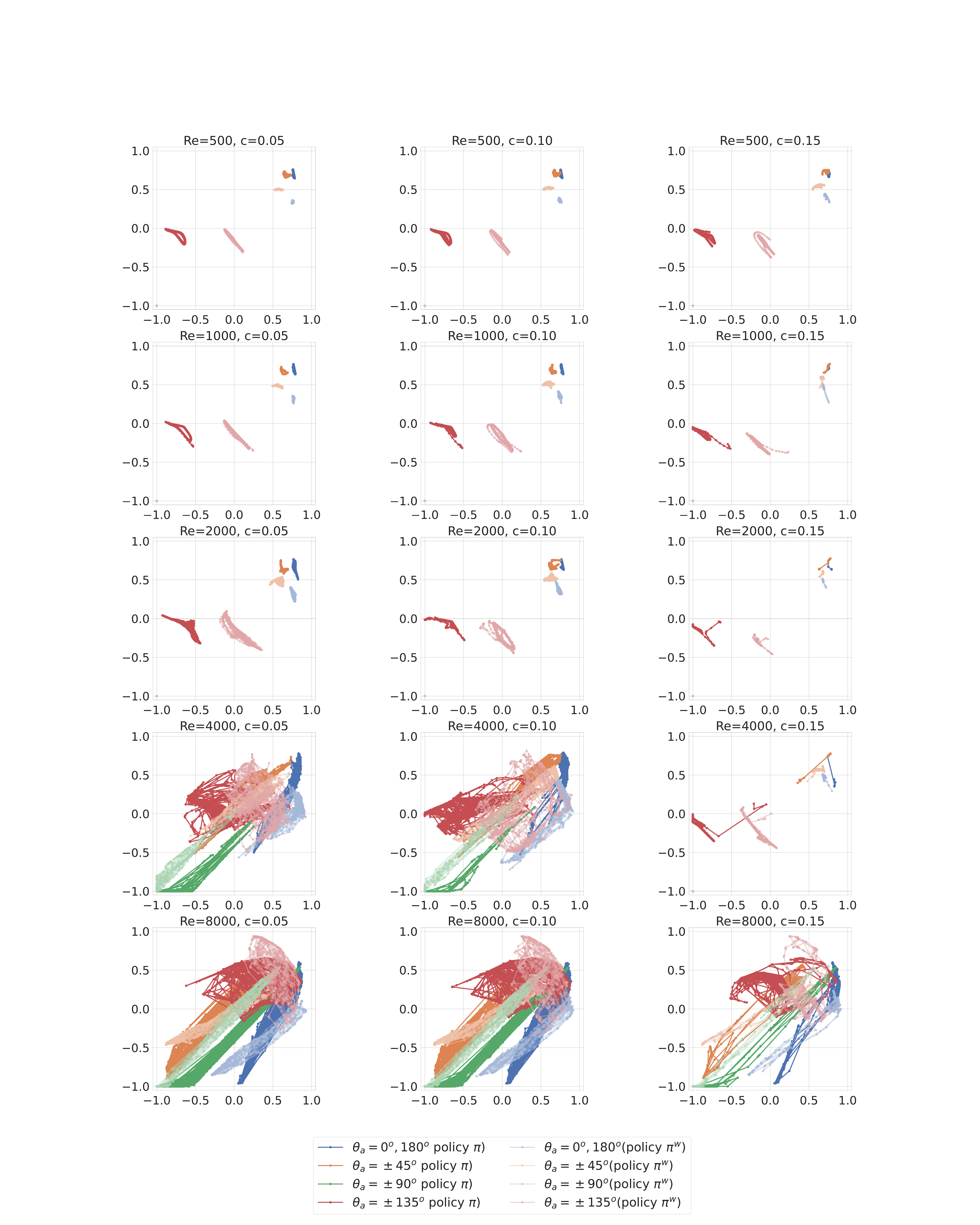}
    \caption{Action pairs  across Reynolds numbers and maximum actuation velocities for the two policies $\pi$ and $\pi^w$. For each plot, the horizontal axis corresponds to $\theta_a \in \{0^o, 45^o, 90^o, 135^o\}$ and the vertical axis to $\theta_a \in \{-135^o, -90^o, -45^o, 180^o\}$.}
    \label{fig:actionsummary}
\end{figure}

Reinforcement learning policies can be effective but at the same time they are complex and do not readily render themselves to interpretation. Here, in order to enhance our understanding of the discoevered policies  we examine their action space. For each Reynolds number, value of $c$, and computed policy ($\pi$ and $\pi^w$) we plot four pairs of actions; each pair corresponds to all actions taken by two actuators, plotted against one another. This is shown in \cref{fig:actionsummary}.

At low Reynolds numbers, we observe mostly constant actions. However, as the Reynolds number increases, more complex behavior emerges. At higher Reynolds numbers, we even observe actuator pairs that switch from blowing to suction.

A similar trend can be observed as the maximum actuation velocity ($c$) decreases. Larger values of $c$ demonstrate a more simplistic approach to drag reduction, where each actuator pair predominantly performs either suction or blowing, without significant variance. On the other hand, reducing the actuation velocity calls for a more sophisticated approach, resulting in increased variance in the actions taken.

\Cref{fig:actionsummary} also demonstrates how introducing a regularizer affects the policy and the action space. In most cases, the trajectories plotted are shifted closer to the origin, indicating a reduction in the action magnitudes for actuators that do not contribute significantly to drag reduction, such as those placed at $\theta_a = \pm 135^{\circ}$. Conversely, the other pairs of actuators, whose actuation velocity magnitude is not significantly reduced by the regularizer, display a strong correlation. This correlation is expected due to symmetry; there is no apparent reason for our policy to prefer one direction over another, especially for the key locations at $\theta_a = \pm 45^{\circ}$ and $\theta_a = \pm 90^{\circ}$, which strongly influence flow separation.

\clearpage
\clearpage

\subsection{Three-dimensional flow}\label{sec:three dimensions}
\begin{table}[!b]
\centering
\caption{Summary of the three-dimensional controlled and uncontrolled cases.}\label{tab:summary-3D}
\begin{ruledtabular}
\begin{tabular}{ccc}
 Re     & $C_D$  & $C_D$, policy $\pi$ ($\%$ reduction)  \\ \midrule
\toprule
 1000   &  0.882              &  0.757 ($14.1\%$) \\
 2000   &  0.916              &  0.750 ($18.2\%$)  \\ 
 4000   &  0.940              &  0.788 ($16.1\%$) \\ 
\end{tabular}
\end{ruledtabular}
\end{table}

\begin{figure}[!b]
    \centering
    \includegraphics[width=\linewidth]{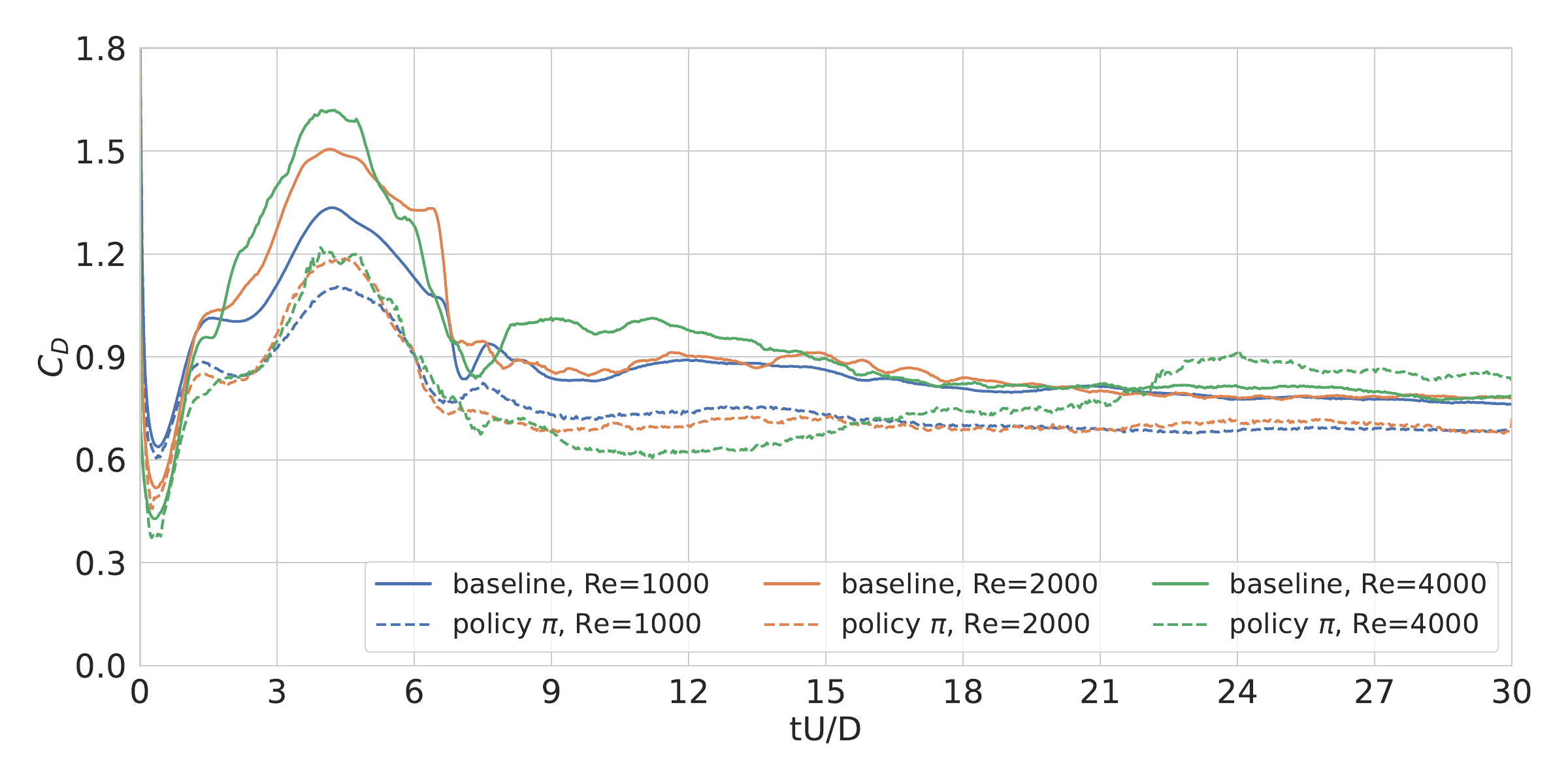}
    \caption{Flow past three-dimensional cylinder at $Re=1000,2000$ and $4000$. Drag coefficient time series for baseline case and for three different policies.}
    \label{fig:3ddrag}
\end{figure}

\begin{figure}[!b]
    \centering
    \includegraphics[width=.48\linewidth]{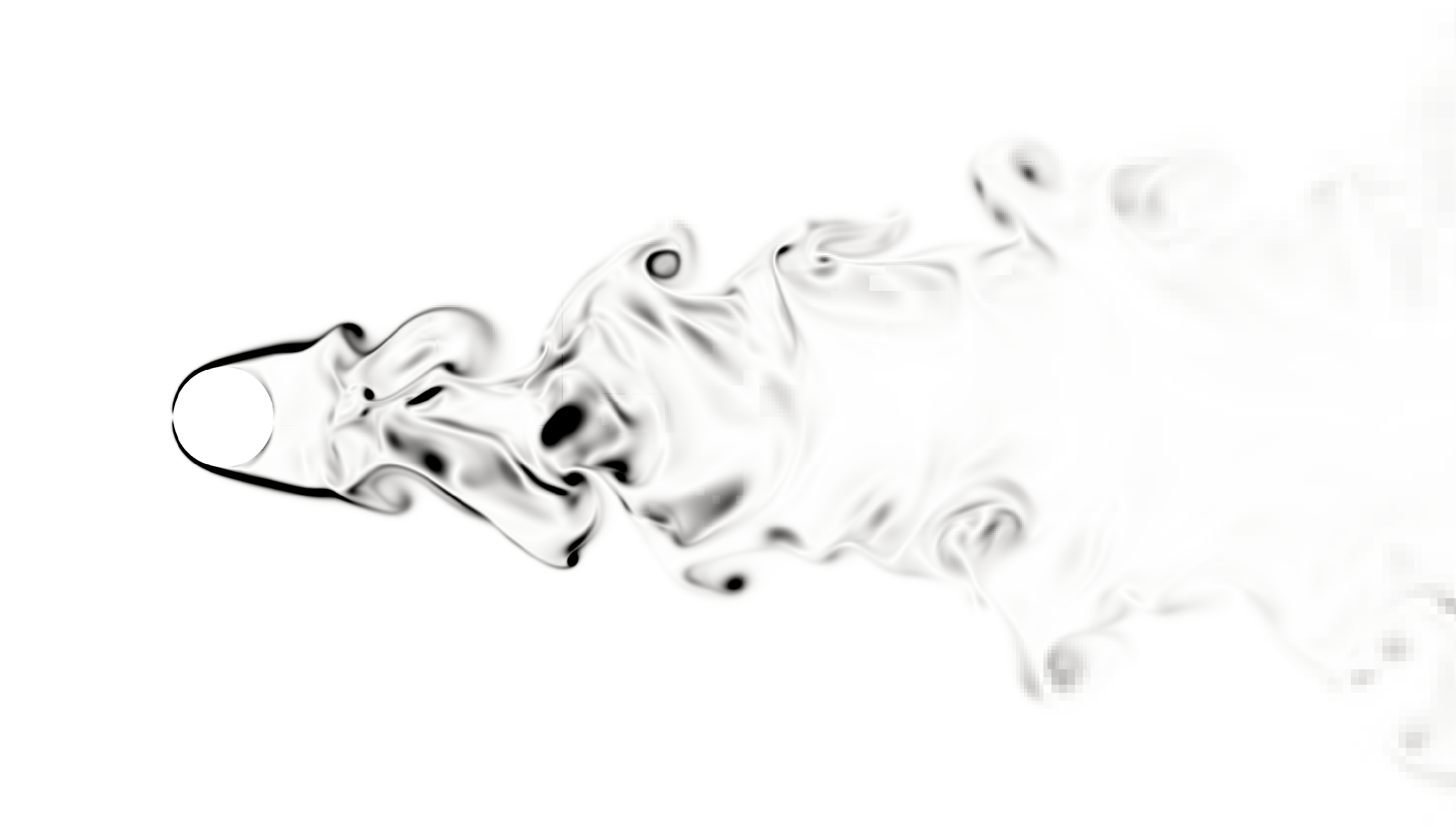}
    \includegraphics[width=.48\linewidth]{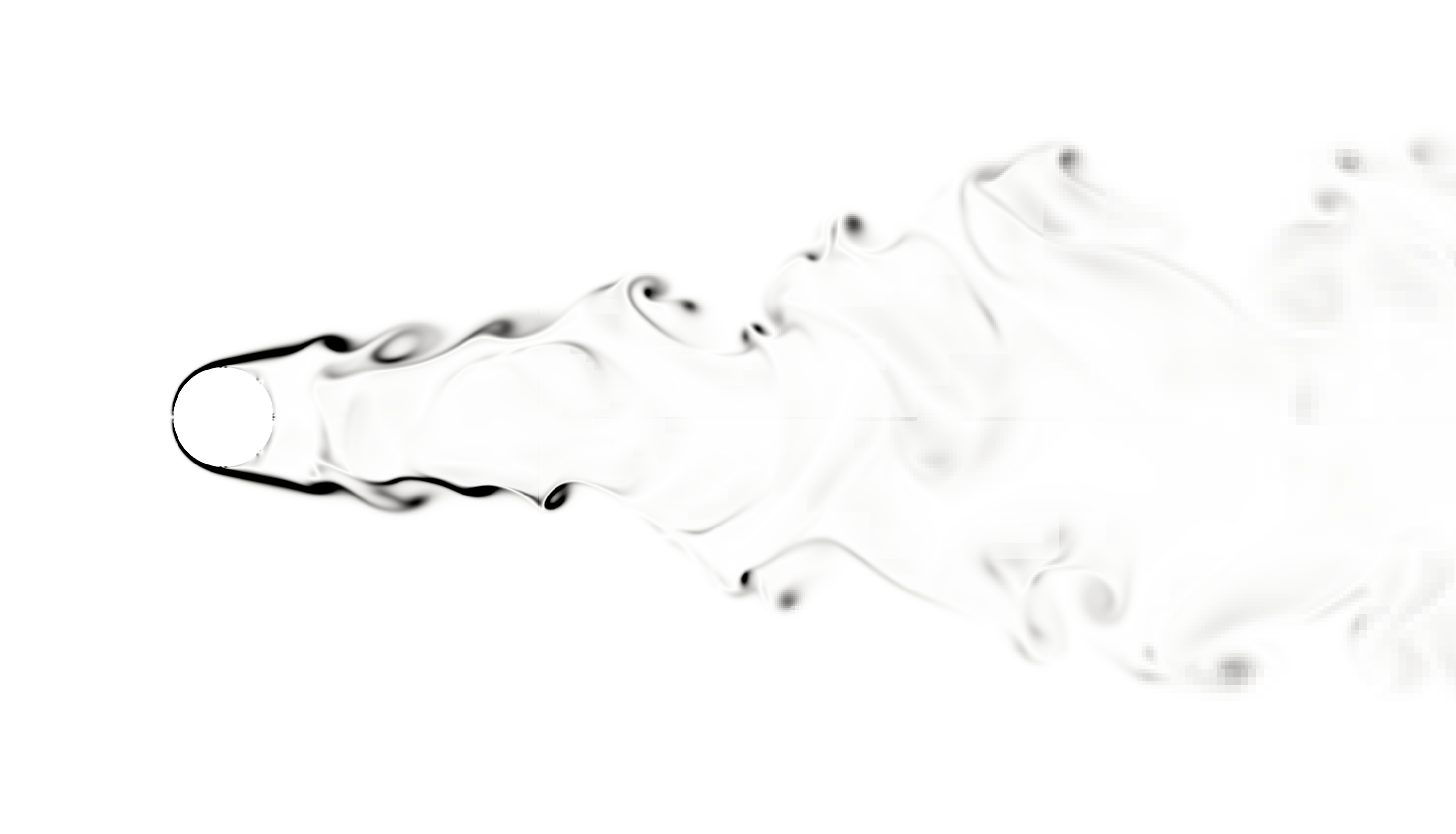}
    \includegraphics[width=.48\linewidth]{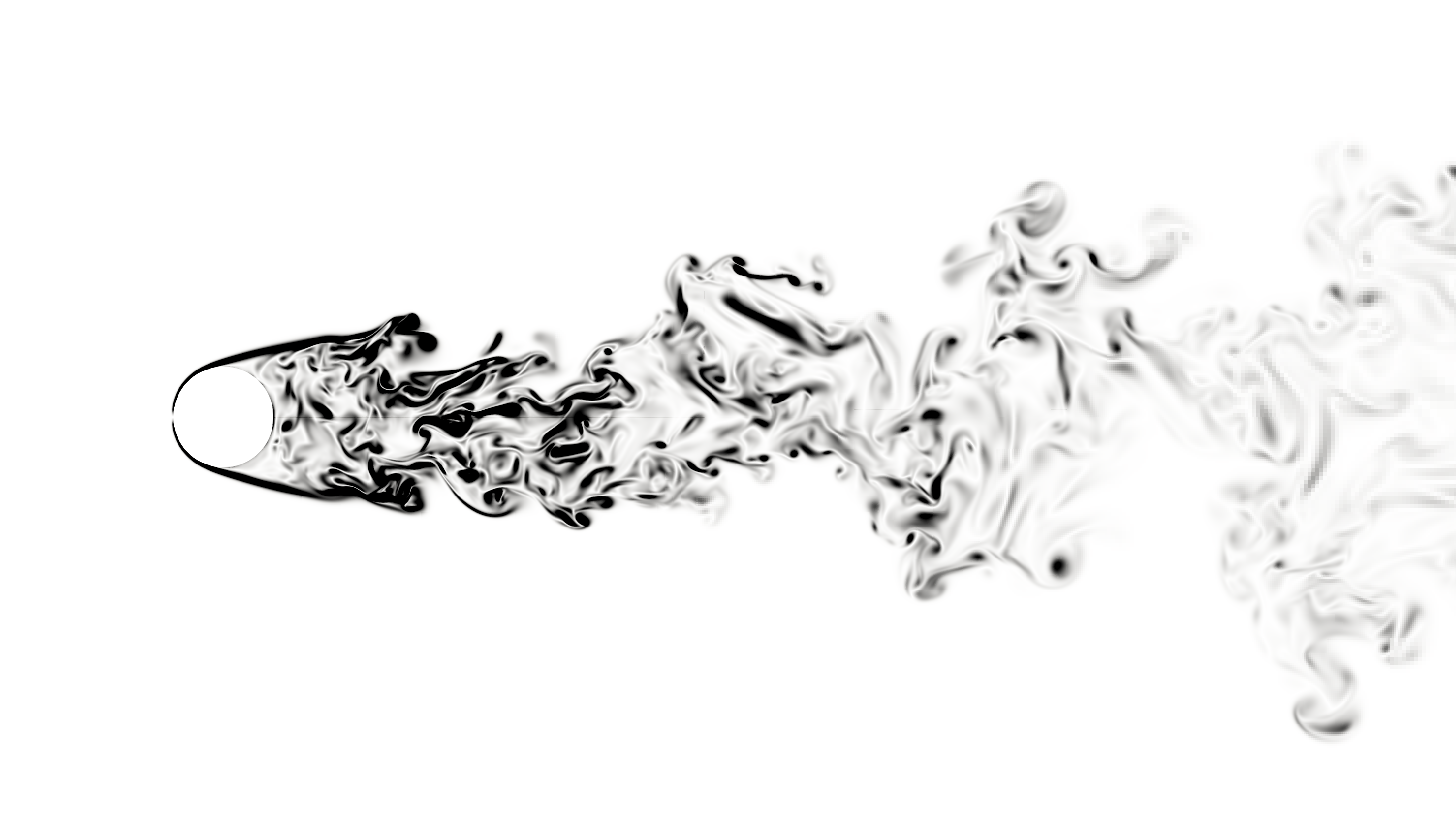}
    \includegraphics[width=.48\linewidth]{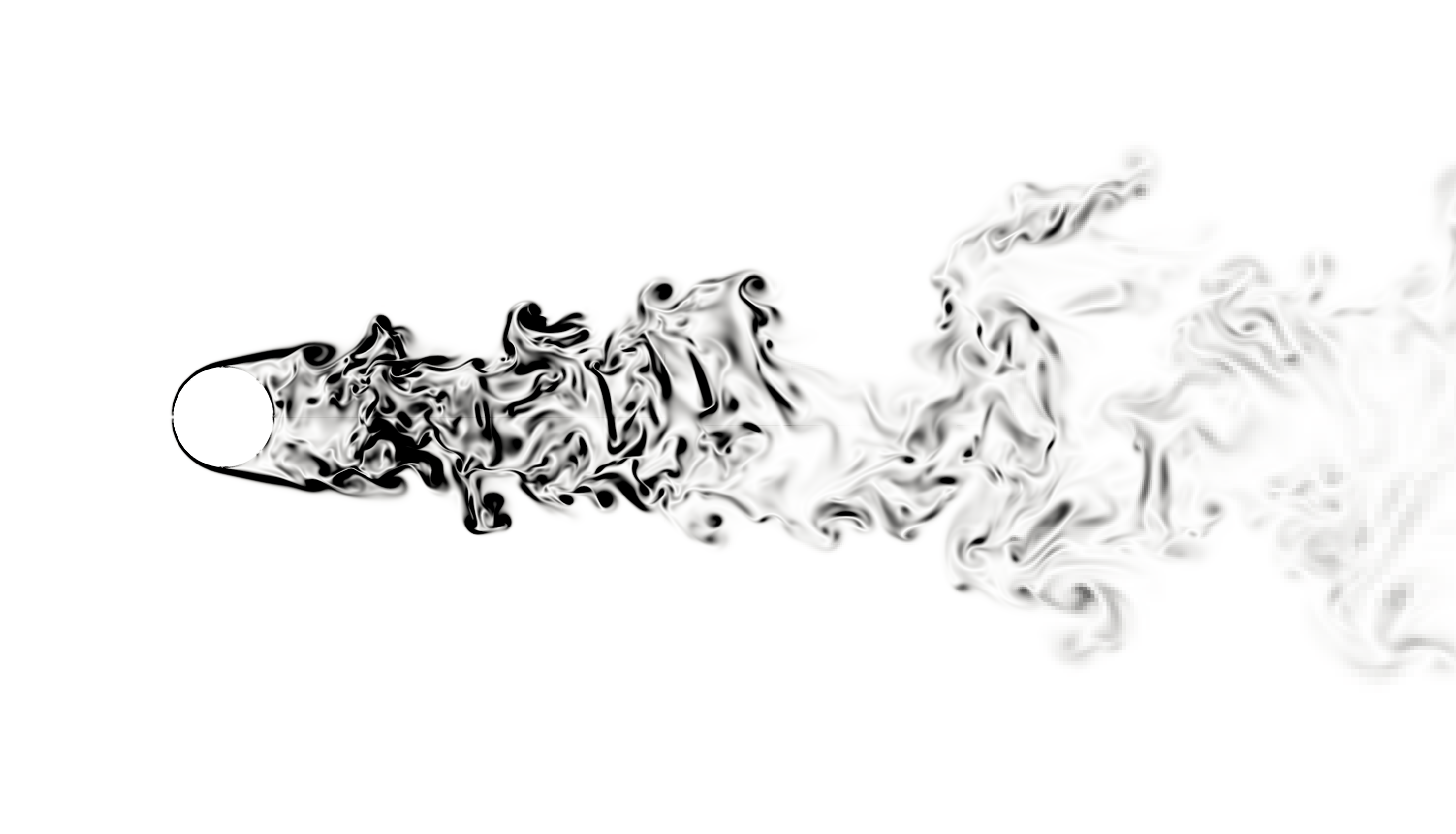}
    \includegraphics[width=.48\linewidth]{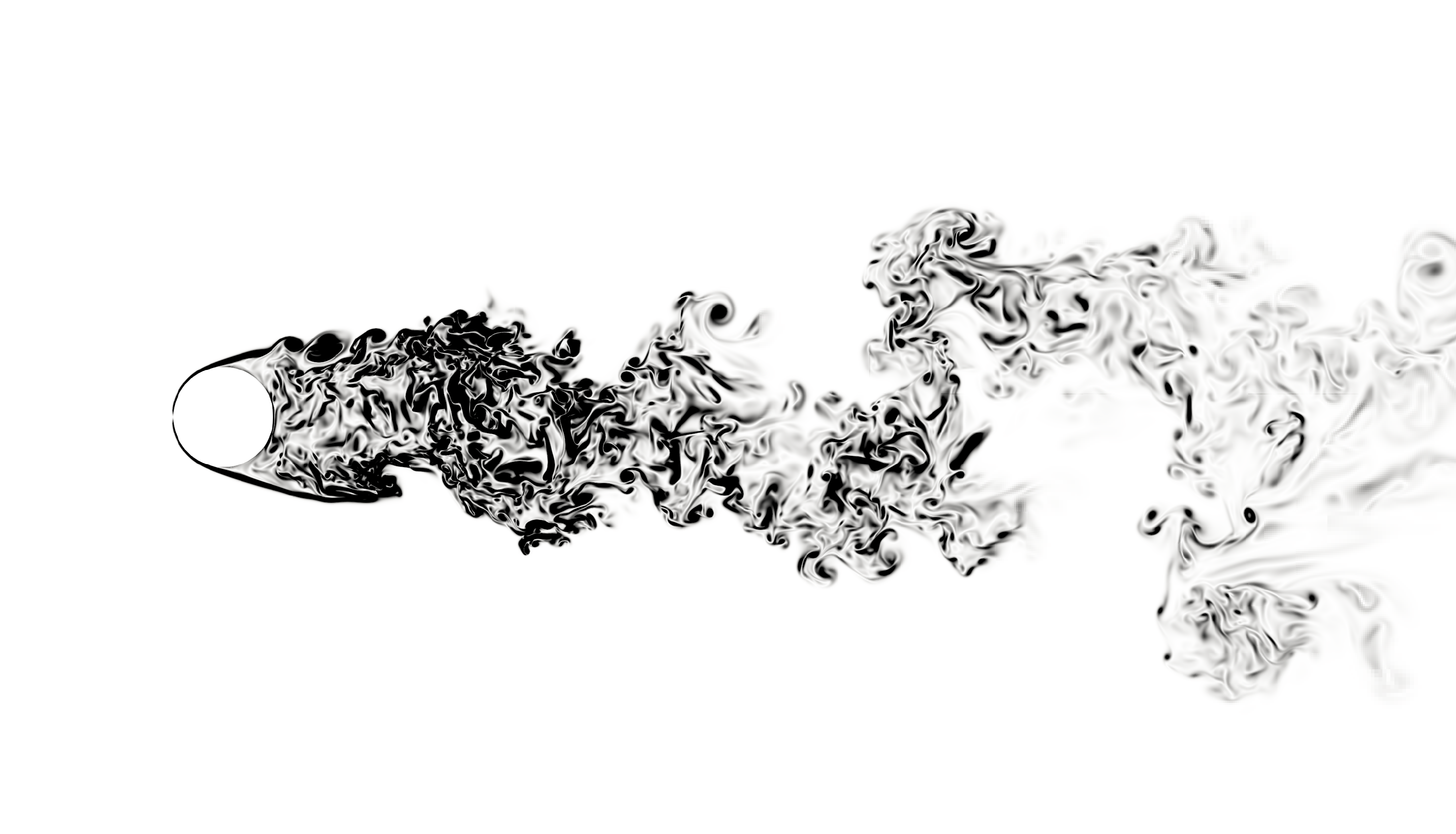}
    \includegraphics[width=.48\linewidth]{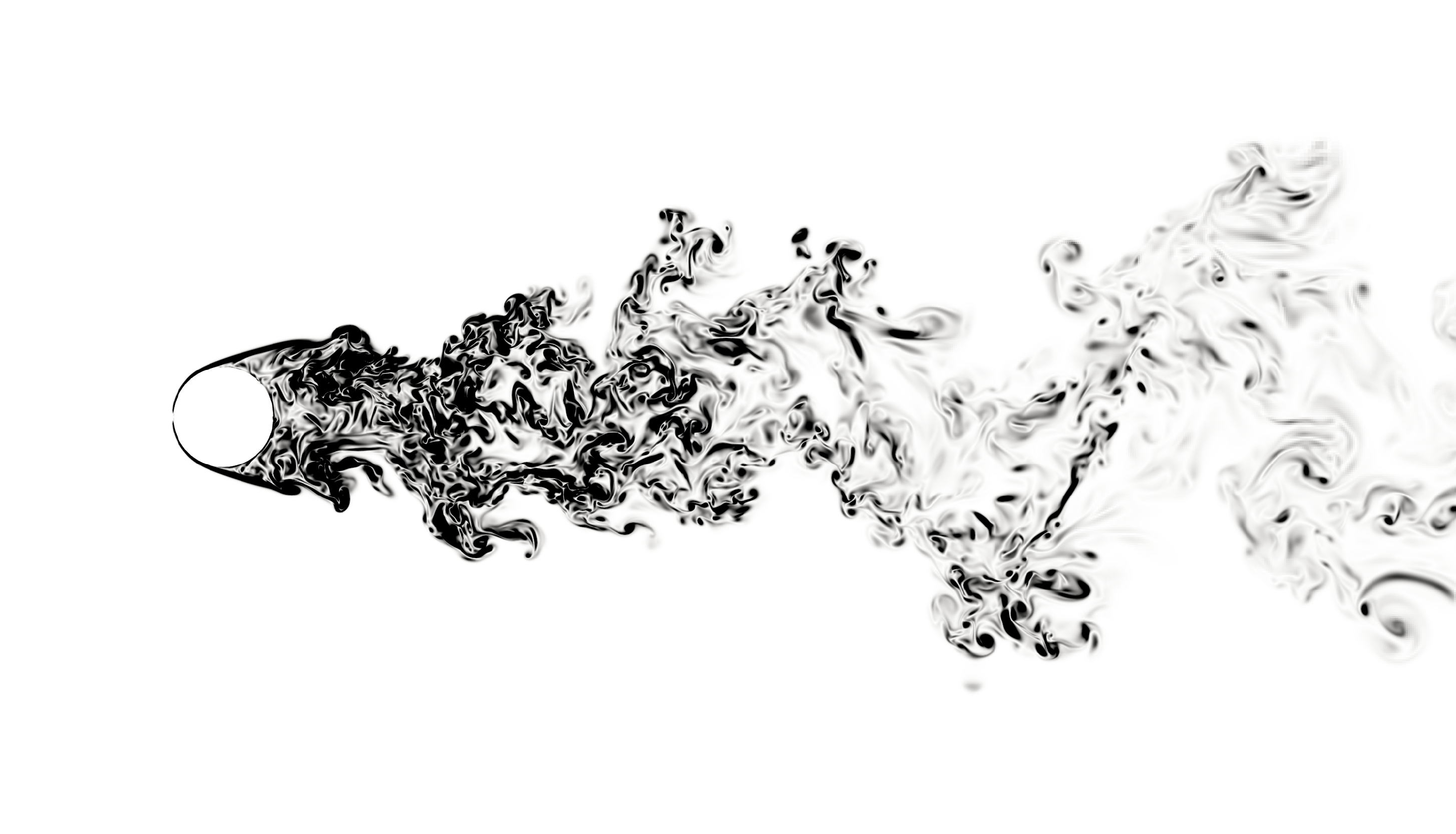}
    \caption{Left column: slice through the xy-plane showing vorticity magnitude at $T=30$, for the uncontrolled cases (top to bottom: $Re=1000,2000,4000$).
    Right column: slice through the xy-plane showing vorticity magnitude at $T=30$, for the controlled cases (top to bottom: $Re=1000,2000,4000$).}
    \label{fig:3dvort}
\end{figure}
 The transition from two-dimensional to three-dimensional vortex shedding has been observed at Reynolds numbers as low as $200$ \cite{Kanaris2011}. Yet, two-dimensional simulations could provide insights relevant to their three-dimensional counterparts at a greatly reduced computational cost; each three-dimensional simulation utilizes about half of the computational resources required to perform a complete training for a single policy (with thousands of two-dimensional simulations). 

Hence, direct training of a three-dimensional model at the range of Reynolds numbers considered in this study is currently not feasible. Nevertheless, we can assess the performance of our policies in a three-dimensional setting. To this end, we apply our computed policy $\pi$ directly to the three-dimensional flow past a cylinder at $Re=1000$, $2000$ and $4000$, for a maximum actuation velocity up to $15\%$ of the cylinder velocity ($c=0.15$).

To implement this, we define a state and a set of actions. Similar to the two-dimensional model, we place $N_a=8$ mass transpiration actuators on the cylinder surface, but in this case, the actuators are extended in the $z$-direction on the cylinder surface, for $|z|<0.5L$. Our choice of state involves sampling pressure and the $z-$component of vorticity at 16 locations on the cylinder surface, with the quantities averaged in the $z$-direction. Additionally, we include the cylinder lift and drag coefficients in our state, as well as the Reynolds number and maximum actuation velocity ($c$). The actuator activation time is set to $T=0.1$ (instead of $T=200$), and the simulations are terminated at $T=30$ (instead of $T=300$). The setup is illustrated in \cref{fig:simulation_setup}.

\begin{figure}[!b]
    \centering
    \includegraphics[width=.48\linewidth]{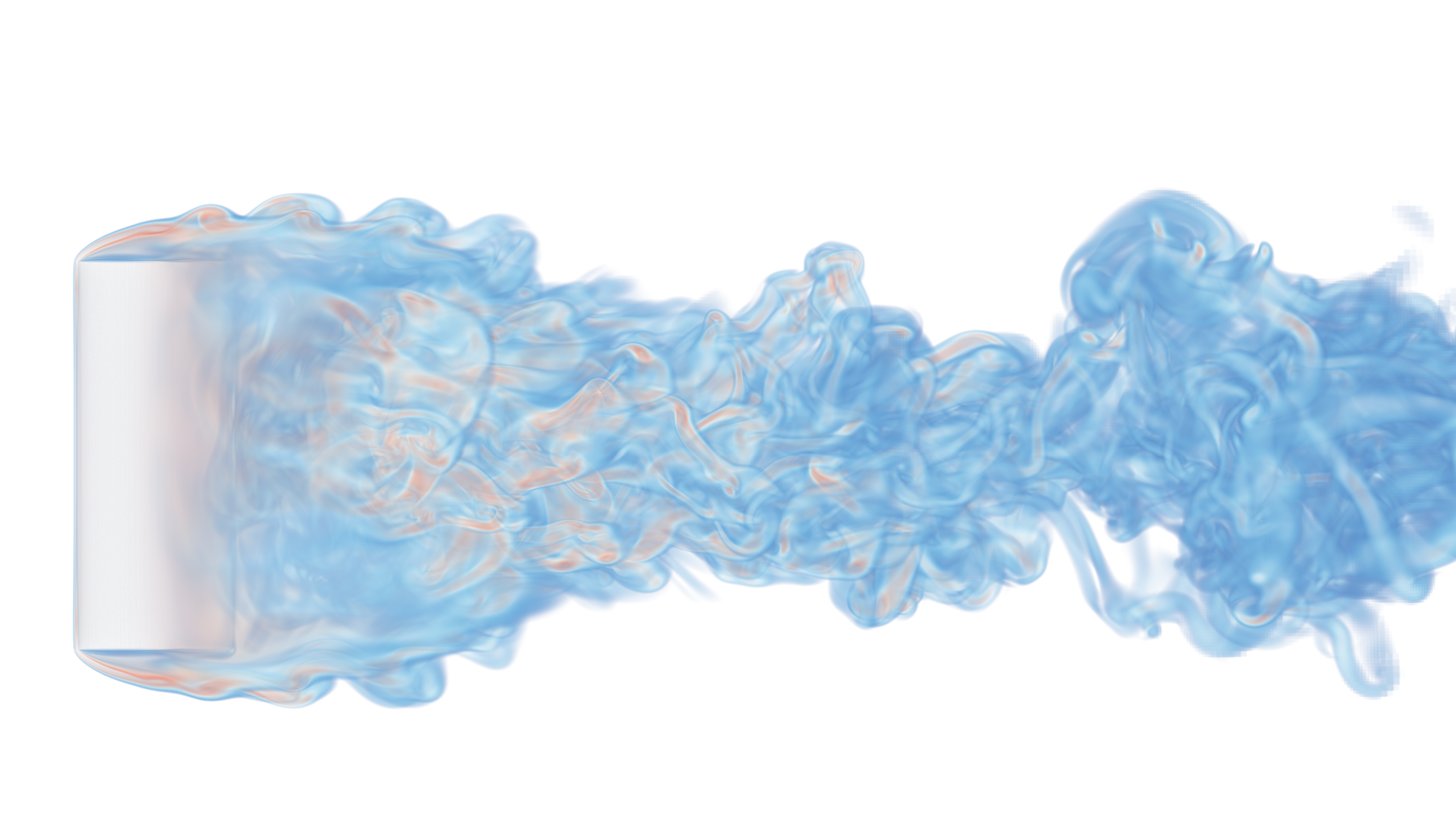}
    \includegraphics[width=.48\linewidth]{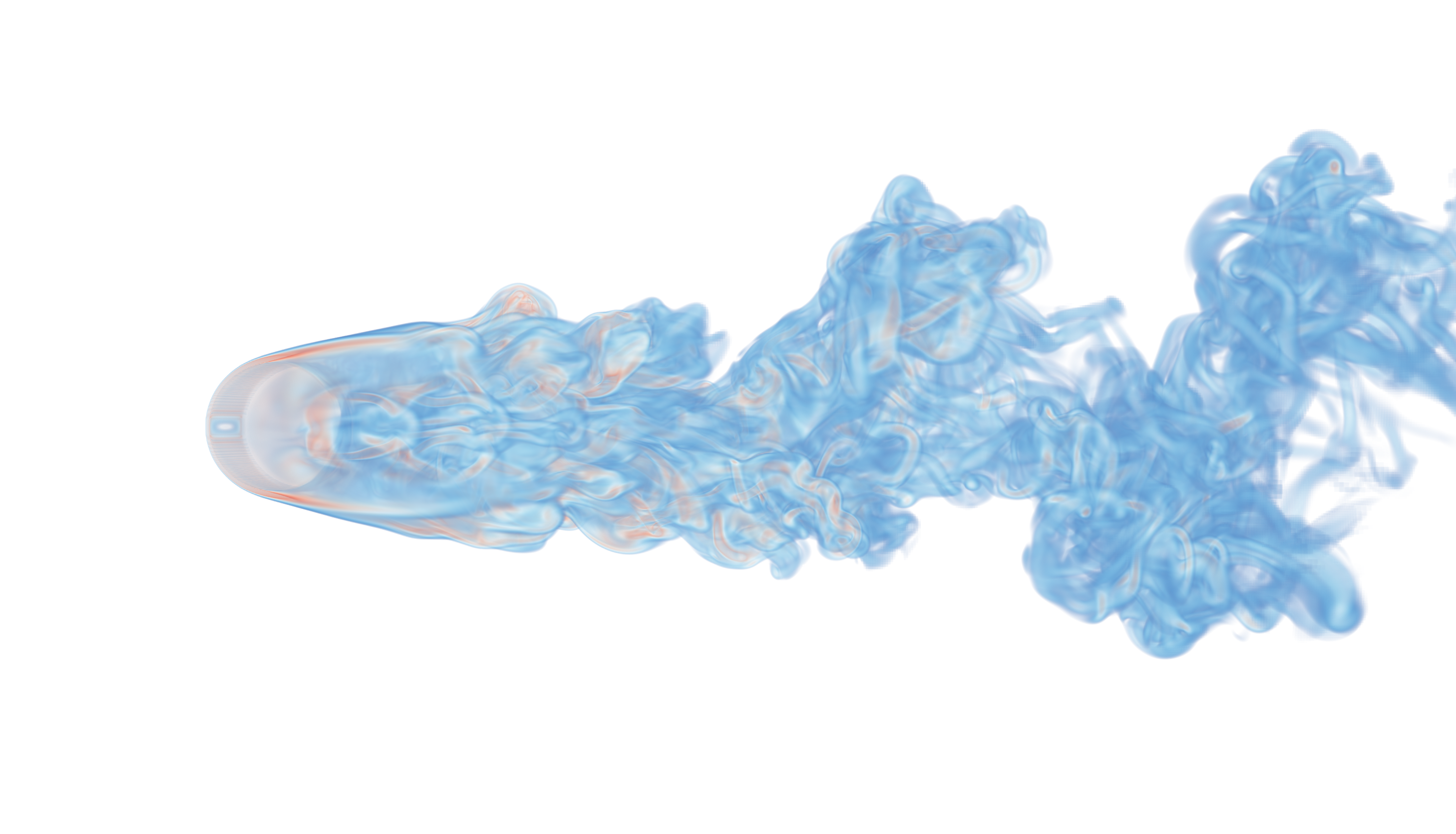}
    \includegraphics[width=.48\linewidth]{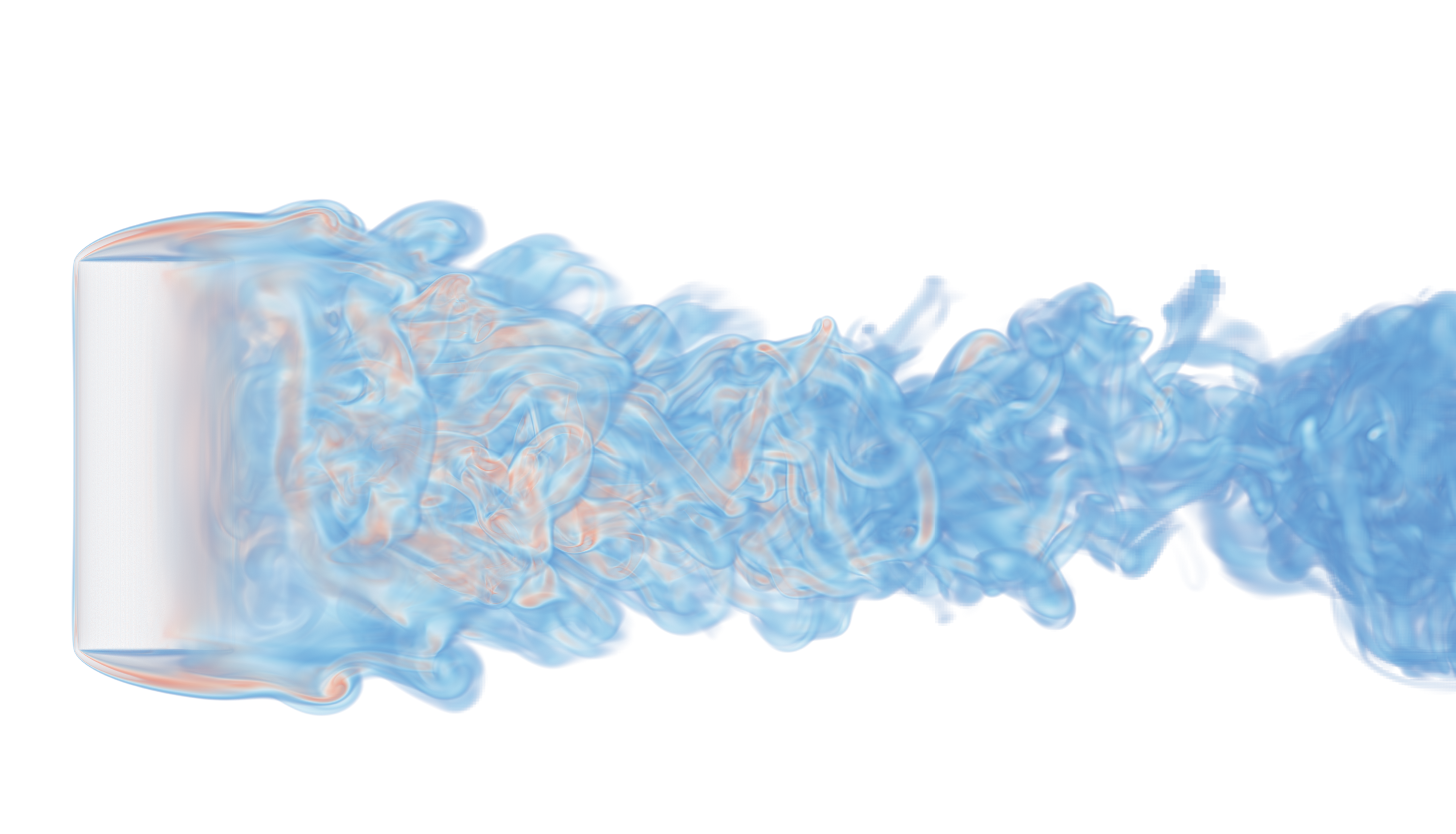}
    \includegraphics[width=.48\linewidth]{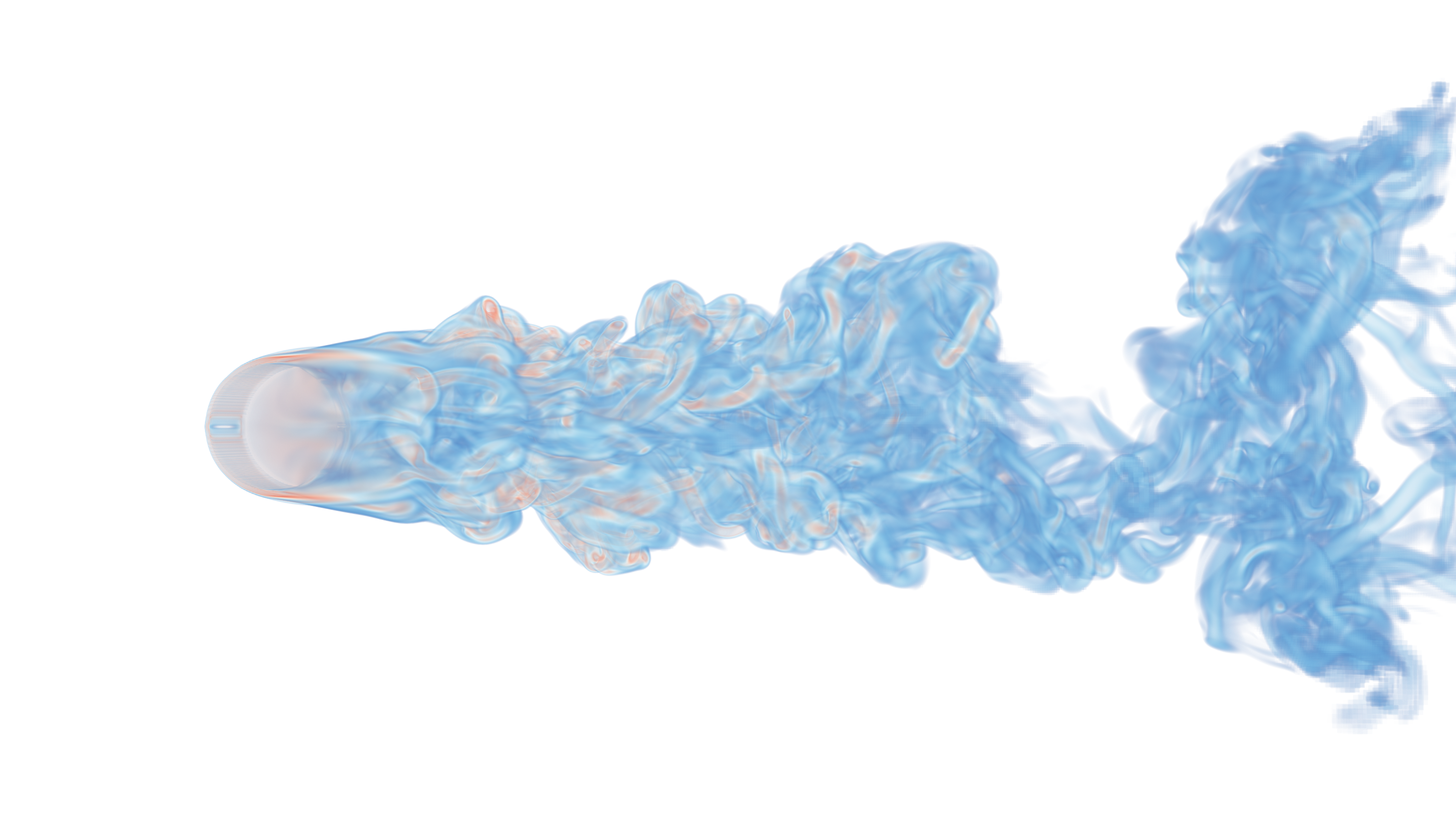}
    \caption{Vorticity magnitude volume rendering for $Re=2000$, top and side views. First row shows the uncontrolled case, second row shows the controlled case.}
    \label{fig:3drender}
\end{figure}
The resulting drag coefficient for all three cases is plotted in \cref{fig:3ddrag} and the results are summarized in~\cref{tab:summary-3D}. The  policy successfully reduces drag, on average, for all three cases. However, the drag we obtain by applying the computed policy is not strictly lower than the baseline case, for $Re=4000$. At about $T=22$ drag starts to increase. Upon closer examination of the actions taken by our policy, it becomes apparent that the actuation velocities are significantly reduced during this time period. Consequently, the reduction in flow control strength leads to the observed increase in drag. This discrepancy can be attributed to the divergence between the observed state by our RL agent and its two-dimensional counterpart, particularly at this Reynolds number. Our policy seems to not be trained well in this regime and would require further training with three-dimensional simulations, to further improve. Still, many observations that were valid in the two-dimensional regime also hold here. As can be seen in \cref{fig:3dvort}, the wake for the controlled cases at $Re=1000$ and $2000$ is narrower and separation is delayed. The same is true for $Re=4000$, but only before $T=22$. Finally, a volume rendering of the vorticity magnitude at $T=30$ for $Re=2000$ is shown in \cref{fig:3drender}, for both the uncontrolled and the controlled cases. 

\section{Conclusions}\label{sec:conclusions}
We investigate drag reduction for 2D and 3D cylinder flows through active control mechanisms discovered by Deep Reinforcement Learning. The study reveals an intriguing trade-off between drag reduction and actuation energy expenditure. Aggressive actuation leads to significant delays  in separation on the cylinder surface and up to $35\%$ drag reduction. On the other hand, a more conservative  approach in terms of energy expenditure,identifies instances when the actuators can be turned off. Similarly, when limited resources are available (in terms of maximum actuation velocity), the discovered policies manipulate the flow field and eventually reduce drag, by  turning on and off, at select time instances, the actuators.

We find that the identified control policies exhibit generalization to a wider range of Reynolds numbers than the ones used during training. Notably, despite being trained for two-dimensional planar flows, the computed policy is effective for three-dimensional flows as well.
We also describe efforts to interpret the complex policies developed by reinforcement learning. At the same time we note that reonfprcement learning requires length evaluations and there is significant room for improvement by developing effective surrogate models. The exhibited generalisation from 2D to 3D flows in intriguing and deserves further examination. We argue that further work in interpretable reinforcement learning is required.

The present paper paves the way for identifying effective and interpretable control strategies while promoting  efficient resource utilization.  We believe that the proposed reinforcement learning strategies for flow control can be extended to a broader range of unsteady separated flows, providing new insights into the drag reduction mechanisms under energy and other constraints.


\bibliography{Bibliography}
\end{document}